\documentclass[12pt]{article}
\usepackage{jheppub}

\def\CN{{\cal N}}

\def\CO{{\cal O}}
\def\IC {{\mathbb C}}
\def\IR {{\mathbb R}}

\def\IZ {{\mathbb Z}}

\title{Surface defects and instanton partition functions}

\author[1]{Davide Gaiotto and Hee-Cheol Kim}

\affiliation[1]{Perimeter Institute for Theoretical Physics\\%
31 Caroline Street North, ON N2L 2Y5, Canada}

\abstract{We study the superconformal index of five-dimensional SCFTs and the sphere partition function of four-dimensional gauge theories 
with eight supercharges in the presence of co-dimension two half-BPS defects. 
We derive a prescription which is valid for defects which can be given a ``vortex construction'', i.e. can be defined by RG flow from vortex configurations in a larger theory. 
We test the prescription against known results and expected dualities. We employ our prescription to develop a general computational strategy 
for defects defined by coupling the bulk degrees of freedom to a Gauged Linear Sigma Model living in co-dimension two.}

\begin{document}
\maketitle
\section{Introduction}
Supersymmetric gauge theories with eight supercharges admit half-BPS co-dimension two defects which 
can play a very useful role in uncovering hidden structures in the bulk theory 
\cite{Gukov:2006jk,Gomis:2007fi,Alday:2009fs,Alday:2010vg,Gaiotto:2012xa}: many protected structures associated to
lower-dimensional theories with four supercharges can be extended to these half-BPS defects and used as a probe of the 
bulk theory with eight supercharges \cite{Gaiotto:2009fs,Gaiotto:2011tf,Gaiotto:2012rg,Gaiotto:2013sma}. 

Canonical examples are half-BPS surface defects in four-dimensional $\CN=2$ gauge theories\footnote{Half-BPS co-dimension two defects also exist for theories with four supercharges, which are much less understood.  Although this paper mainly focuses on theories with eight supercharges, we will spend some time reviewing some 
examples with less supersymmetry which were first analyzed in \cite{Chen:2014rca}.}: 
the very existence of a Seiberg-Witten curve for a theory can be tied to the existence of a 
one-parameter surface defect in the gauge theory and its twisted chiral ring \cite{Gaiotto:2009fs}. 
Surface defects play an important role in computations of the BPS spectrum of the bulk theory \cite{Gaiotto:2011tf,Gaiotto:2012rg}, 
in the AGT correspondence \cite{Alday:2009fs,Alday:2010vg} and in the calculation
of the supersymmetric index for non-Lagrangian theories \cite{Gaiotto:2012xa}.

There are four general strategies to produce surface defects in some supersymmetric gauge theory $T$. 
\begin{enumerate}
\item A modification of the gauge theory path integral, in a manner akin to 'tHooft operators: one imposes that the fields should approach a 
specific singular behaviour at the location of the defect \cite{Gukov:2006jk}. We will denote these as Gukov-Witten defects. 
\item The addition of extra degrees of freedom localized at the defect: for example, 4d gauge fields can gauge some flavor symmetries of a 2d theory with $(2,2)$ supersymmetry, 
and 4d hypermultiplets can enter 2d superpotential couplings. 
\item Renormalization Group flow from vortex-like configurations: given a second theory $T'$ with an RG flow to $T$ triggered by a scalar operator branch vev,
it is possible to define surface defects in $T$ in terms of a position-dependent vev (``vortex'') in $T'$ \cite{Gaiotto:2012xa}. 
\item In the context of a brane engineering construction, some surface defects can be engineered by adding extra D-branes to the system \cite{Alday:2009fs}. 
\end{enumerate}
These strategies can be combined, say by coupling local matter to the subgroup of the gauge group 
left unbroken by a GW defect, or by looking at vortices in $T'$ in the presence of a 
surface defect in $T'$. 

The same surface defect may admit multiple dual realizations (at least up to $D$-terms). 
For example, the ``vortex'' construction is expected to produce the same result as 
coupling the four-dimensional gauge fields to a 2d sigma model whose target is a vortex moduli space. 
These dualities can be rather useful, as different constructions may allow different computational strategies. 
String theory constructions and dualities often provide a link between different-looking gauge theory constructions: the vortex construction is the field-theory 
version of ``geometric transitions'' \cite{Gopakumar:1998ki,Aganagic:2011sg}, and brane systems often provide a GLSM description of field theory defects \cite{Hanany:1997vm}.

The supersymmetric index \cite{Kinney:2005ej,Romelsberger:2005eg} (aka $S^1 \times S^3$ partition function) of four-dimensional gauge theories is an example of protected quantity 
which can be enriched by co-dimension two BPS defects (lying on $S^1 \times S^1$), even for gauge theories with four supercharges only. 
\footnote{Similar considerations apply to 
three-dimensional gauge theories with the same amount of supersymmetry, simply by dimensional reduction from the $S^1 \times S^3$ partition function
to the ellipsoid $S^3_b$ partition function. }
There are specific prescriptions to compute the index enriched by BPS operators described either 
by coupling to a 2d gauged linear sigma model \cite{Gadde:2013dda} or by position-dependent scalar vevs in a larger theory $T'$ \cite{Gaiotto:2012xa}. 
In the former case, one essentially inserts an equivariant elliptic genus of the 2d degrees of freedom 
into the index calculation of the four-dimensional theory. In the latter case, one picks certain residues of the 
index of $T'$ . The results are compatible with the expected dualities. 

We expect that similar considerations should apply to a second class of protected quantities, based on equivariant localization 
and instanton moduli spaces: the instanton partition functions \cite{Nekrasov:2002qd,Nekrasov:2003rj} and $S^4$ partition functions \cite{Pestun:2007rz} of 
four-dimensional $\CN=2$ gauge theories and the equivariant index on $R^4$ and on $S^4$ \cite{Nekrasov:2002qd,Kim:2012gu} 
of five-dimensional $\CN=1$ gauge theories. These examples are the main focus of this paper. \footnote{
Further extension to the $S^1 \times S^2$ partition functions of 3d theories and to other supersymmetric partition functions 
in various dimensions should also be possible.}

Currently, localization calculations have been mostly focused on GW-type co-dimension two defects.
The corresponding enriched partition functions are computed by replacing the standard instanton moduli spaces with 
moduli spaces of ``ramified instantons'' \cite{Alday:2010vg}. 

Defects defined by coupling the bulk degrees of freedom to a GLSM in co-dimension two also seem obvious targets for localization computations,
some hybrid of calculations on $S^4$ \cite{Nekrasov:2002qd,Pestun:2007rz,Kim:2012gu} and on $S^2$ \cite{Kapustin:2011jm,Doroud:2012xw,Benini:2012ui}.
Indeed, the localization analysis of \cite{JOEL} indicates that such a calculation is possible, and the only new ingredient required is the 
instanton partition function/equivariant index of the bulk theory on $R^4$, coupled to chiral multiplets living in a co-dimension two plane in $R^4$. 

The full partition function or index should take the familiar form of an integral over Coulomb branch parameters of bulk and defect gauge fields
of an integrand built from perturbative 1-loop factors and instanton contributions from the North and South pole of the sphere. 
For example, an index on $S^4$ with a defect along $S^2$ should take a schematic form 
\begin{equation}
\sum_{m_{\mathrm{3d}}} \oint d\zeta_{\mathrm{3d}} \oint d \alpha_{\mathrm{5d}} Z_{\mathrm{1-loop}}^{\mathrm{5d}}(\alpha_{\mathrm{5d}}) 
Z_{\mathrm{1-loop}}^{\mathrm{3d}}(m_{\mathrm{3d}},\zeta_{\mathrm{3d}},\alpha_{\mathrm{5d}})|Z^{\mathrm{5d/3d}}_{inst}(m_{\mathrm{3d}},\zeta_{\mathrm{3d}},\alpha_{\mathrm{5d}})|^2
\end{equation}
where $\zeta_{\mathrm{3d}}$ and $m_{\mathrm{3d}}$  are the equivariant parameters and $S^2$ magnetic fluxes for the 3d gauge fields 
and $\alpha_{\mathrm{5d}}$ are the equivariant parameters for the 5d gauge fields. The one-loop factors should be the same as for decoupled 
$3d$ and $5d$ systems, up to replacing some 3d flavor symmetry equivariant parameters with appropriate 5d gauge equivariant parameters. 
The crucial instanton partition function/equivariant index $Z^{\mathrm{5d/3d}}_{inst}(m_{\mathrm{3d}},\zeta_{\mathrm{3d}},\alpha_{\mathrm{5d}})$
should only depend on the choice of which 3d chiral multiplets couple to the 5d gauge fields, not on the choice of 3d gauge group. 

The first objective of this paper is to identify the correct prescription to compute instanton partition functions enriched by defects 
defined by a vortex construction. Essentially, the enriched partition function is computed by 
specializing the equivariant parameters of the ``bare'' partition function of $T'$ to special values where a 
(position-dependent) Higgs branch opens up. Similar calculations have appeared before in the literature, but usually specialized to the case where 
the IR bulk theory $T$ is trivial, and thus the specialization of the UV $T'$ partition function produces the partition function of 2d or 3d theories.
See e.g. \cite{Dimofte:2010tz,Dorey:2011pa,Nieri:2013vba,Aganagic:2014oia}. We will test the prescription it by comparison to geometric transitions in 
the topological string literature and from inspection of the AGT correspondence. 

Some of the defects defined by the vortex construction are expected to have a dual description as 
GLSM degrees of freedom coupled to the bulk gauge theory. The second objective of this paper is to compare our results  
with the expected form of localization formulae for GLSM defects. We will successfully match the two prescriptions 
for the simplest example involving three-dimensional Abelian gauge fields: pure five-dimensional $SU(2)$ gauge theory coupled to 
a $CP^{1}$ three-dimensional GLSM. 

\subsection{Structure of the paper}
In section \ref{sec:Higgs} we develop the Higgsing prescription for computing supersymmetric 
indices and partition functions in the presence of co-dimension two defects. As partition functions can be usually obtained 
from dimensional reduction of indices, we first review the prescription for indices of $\CN=1$ and $\CN=2$ four-dimensional gauge theories and 
then extend it to the index of $\CN=1$ five-dimensional gauge theories and thus to the $S^4_b$ partition function of $\CN=2$ four-dimensional gauge theories.

In section \ref{sec:webs} we review the Higgs branch of five-dimensional SCFTs which admit a fivebrane web construction
and identify the corresponding defects with the field theory limit of D3 branes transverse to the fivebrane web. Armed with this geometric 
picture and the Higgsing prescription, we thus proceed to compute the index of several co-dimension two defects in the 5d SCFT 
associated to the pure $SU(2)$ gauge theory in five dimensions. 

We subject our results to a strong check by verifying that the resulting 
indices are related by Witten's $SL(2,\IZ)$ action on 3d SCFTs equipped with a $U(1)$ flavor symmetry. 
In particular, that is a check that the index transforms as expected under gauging a 3d Abelian flavor symmetry. 
In the process of doing this check, we learn how to add background magnetic flux to the 3d/5d index obtained from the Higgsing prescription. 

In section \ref{sec:diff} we further refine the index calculation by inserting Wilson loop operators. 
This allows us to subject our 3d/5d indices to a second stringent test: they satisfy difference equations which 
quantize the Seiberg-Witten curve of the 5d gauge theory. 

In section \ref{sec:equiv} we discuss how to introduce 3d chiral degrees of freedom directly in the calculation of an equivariant index. 

Finally, in section \ref{sec:conc} we present our conclusions and discuss some open questions.

\section{The Higgsing prescription} \label{sec:Higgs}
The superconformal index for four-dimensional $\CN=1$ gauge theories is invariant under RG flows, as long as one can match the 
$U(1)_R$ symmetry generators in the UV and IR. A special application of this principle 
is possible when we have some superconformal theory $T^{UV}$ which has a moduli space of vacua, 
parameterized by the vev of some collection of chiral operators $\CO_i$: we can initiate an RG flow by the operator vevs and 
attempt to match the superconformal index of $T^{UV}$ to the index of whatever degrees of freedom $T_{IR}$ which can be found at the end of the RG flow. 

This match is possible only if we can find a linear combination $f'_R = f_R + c^a f_a$ of the $U(1)_R$ symmetry generator $f_R$ 
and other flavor symmetry generators $f_a$ of the theory which is preserved by the $\CO_i$ vevs, which we can use to define a
new $U(1)'_R$ symmetry unbroken along the flow. If the $\CO_i$ have R-charge $q_R^i$ and flavor charges $q_a^i$, 
that means we need $q_R^i = - c^a q_a^i$. In particular, the $\CO_i$ must all carry flavor symmetry. 

There is a close relation between the existence of such directions in the vacuum moduli space
and the poles of the superconformal index of $T^{UV}$ as a function of the flavor fugacities. 
Each time we can find a dimension $d$ sub-manifold in the space of vacua where the only chiral operators which get a vev 
have charges proportional to some elementary charge $(q_R,q_a)$, we expect the index to have a pole of order $d$ 
at 
\begin{equation}
(p q)^{\frac{q_R}{2}} \prod_a z_a^{q_a} =1 \ .
\end{equation}

If we consider the index as a partition function on $S^3 \times S^1$, the pole is due to the presence of $d$ 
bosonic zero modes: the constant vevs along the sub-manifold of vacua. 
The leading coefficient of the divergence should be controlled by the physics far along the sub-manifold of vacua, 
i.e. by the index of $T_{IR}$. 

At the end of the RG flow, one usually finds some IR SCFT $T_{IR}$ accompanied by a collection of 
free fields. At the very least, we will have $d$ free chiral fields associated to the fluctuations along the 
sub-manifold of vacua. If the vacua break spontaneously some other non-Abelian symmetry generators, 
we will have some extra Goldstone bosons as well. 

For simplicity, we will restrict ourselves to the case $d=1$, and take the sub-manifold of vacua to be a complex line parameterized 
by a complex parameter of charges $(q_R,q_a)$. In order to find the index of $T_{IR}$ we can take the following operation:
\begin{equation}
I_{IR}(z_a;p,q) = I^{-1}_{\mathrm{free}}(z_a;p,q) \lim_{(p q)^{\frac{q_R}{2}} \prod_a z_a^{q_a} \to 1} I^{-1}_{\mathrm{chiral}}\left((p q)^{\frac{q_R}{2}} \prod_a z_a^{q_a};p,q\right) I_{UV}(z_a;p,q) \ .
\end{equation}
Here the chiral multiplet index $I_{\mathrm{chiral}}((p q)^{\frac{q_R}{2}} \prod_a z_a^{q_a})$ subtracts the contribution from the 
flat direction itself, and eliminates the pole in the UV index $I_{UV}(z_a;p,q)$. We also subtract the contribution $I_{\mathrm{free}}(z_a;p,q)$ of 
other extra free fields appearing in the IR. 

The index of a chiral multiplet is written as 
\begin{equation}
I_{\mathrm{chiral}}(z;p,q) = \Gamma(z;p,q) \equiv \frac{(p q z^{-1};p,q)_\infty}{(z;p,q)_\infty}
\end{equation}  
in a convention where the R-charge of the scalar component is set to $0$. Note the definition 
$(z;p,q)_\infty = \prod_{n,m\geq0} (1-z p^n q^m)$.
In general, a chiral multiplet of R-charge $q_R$ would contribute $I_{\mathrm{chiral}}((pq)^{\frac{q_R}{2}}z;p,q)$.

If we define a fugacity $z = \prod_a z_a^{-q_a}$, we can write the final result as a residue 
\begin{equation}
I_{IR}(z_a;p,q) = I^{-1}_{\mathrm{free}}(z_a;p,q) I'_{\mathrm{gauge}}(p,q)\mathrm{Res}_{z \to (p q)^{\frac{q_R}{2}} } I_{UV}(z_a;p,q) \ .
\end{equation}
Here $I'_{\mathrm{gauge}}(p,q)$ is the index for a free Abelian gauge multiplet with zero mode removed. 

The index will in general have other poles besides the ones associated to constant flat directions on the sphere. 
Each of the poles we have discussed above will typically be accompanied by a doubly-infinite tower of 
poles associated to position-dependent vevs with angular momenta $r$ and $s$ on $S^3$. These poles will be located at 
\begin{equation}
p^r q^s (p q)^{\frac{q_R}{2}} \prod_a z_a^{q_a}=1 \ .
\end{equation}
The position-dependent vev will reduce $T_{UV}$ to $T_{IR}$ almost everywhere, except at the zeroes of the vev. The result is that 
$T_{IR}$ will be modified by co-dimension two defects on the two $S^1$ in $S^3$ fixed by the rotation Cartan generators. 
The defects will be labelled by $r$ and $s$ respectively. 

Thus we can compute the index of $T_{IR}$ modified by the surface defects with label $r$ and $s$ as 
\begin{equation}
I^{r,s}_{IR}(z_a;p,q) = I^{-1}_{\mathrm{free}}(z_a;p,q) \lim_{p^r q^s (p q)^{\frac{q_R}{2}} \prod_a z_a^{q_a} \to 1} I^{-1}_{\mathrm{chiral}}\left((p q)^{\frac{q_R}{2}} \prod_a z_a^{q_a};p,q\right) I_{UV}(z_a;p,q) \ .
\end{equation}
Notice that the $I^{-1}_{\mathrm{chiral}}$ factor inside the limit cancels all the relevant poles in the partition function $I_{UV}(z_a;p,q)$,
and gives us a finite limit. 

For general gauge theories the index is written as a contour integral over gauge fugacities. Poles of the final answer as a function of 
flavor fugacities usually arise from the collision of two or more poles in the integrand pinching off the integration contour of some gauge fugacities. 
Very roughly, the integrand poles are associated to individual chiral fields which receive a vev and the pinched contours to Cartan gauge fields which are being Higgsed. 
Typically, cancellations between chiral multiplet and W-boson contributions will also occur. 
Upon taking the residue at $r=s=0$, we are left with a contour integral which describes the index of the gauge theory which is left upon the Higgsing procedure. 

When taking a general residue, the final answer will arise from a similar pattern of pinching and cancellations. Due to the shifts of arguments of the various factors by 
powers of $p$ and $q$, and relations such as 
\begin{equation}
I_{\mathrm{chiral}}(p z;p,q) = \Gamma(p z;p,q) =  \Gamma(z;p,q) (q z^{-1};q)_\infty (z;q)_\infty \equiv \theta(z;q) \Gamma(z;p,q) \ ,
\end{equation} 
the final answer will typically be a sum of terms involving the same $\Gamma(z;p,q)$ functions as for $r=0$, $s=0$ and various combinations of theta functions,
which can be very roughly understood as elliptic genera of 2d $(0,2)$ fermi and chiral multiplets. 
We will give a few examples in a later section. 

As explained in \cite{Gaiotto:2012xa}, the Higgsing prescription has an alternative interpretation in terms of 
vortices, by weakly gauging the $U(1)$ flavor symmetry we are planning to take a residue by, and 
turning on a large FI parameter to allow for the existence of vortices which induce the desired 
position-dependent vevs of the fields. There is an interesting intermediate point of view, which helps giving a physical meaning to the 
position-dependent vevs without gauging the $U(1)$ flavor symmetry: we can create the low energy 
co-dimension 2 defects by turning on a {\it background vortex} in $T_{UV}$. 
What we mean with that is turning on a background $U(1)$ flavor symmetry connection,
say for example independent of the $x^3$,$x^4$ directions and invariant under rotations in the $x^1$,$x^2$ 
plane, consisting of $r$ units of $U(1)$ flux. 

The theory can still have BPS configurations, 
which solve essentially the same equations as BPS vortices, except for the D-term equation 
for the background $U(1)$ flux. In particular, chiral operator vevs are covariantly holomorphic. 
The background flux allows for the operators to have vevs with zeroes of appropriate order
near the origin, without blowing up at infinity. Such vevs initiate the RG flows which lead to the 
co-dimension two defects discussed above.

It is straightforward to propose similar prescriptions for other protected quantities. 
In the context of theories with four supercharges, we simply replace the $I^{-1}_{\mathrm{chiral}}$ and $I^{-1}_{\mathrm{free}}$ 
factors with the corresponding protected quantities in the appropriate dimension. For 3d $\CN=2$ theories we can consider either the ellipsoid 
partition function~\cite{Kapustin:2012iw,Hama:2011ea} or the superconformal index~\cite{Kim:2009wb,Imamura:2011su}. The former case is a straightforward dimensional reduction of the four-dimensional index~\cite{Dolan:2011rp,Gadde:2011ia}. 
We will use the partition function of a free chiral 
\begin{equation}
Z^b_{\mathrm{chiral}}(\sigma,b) = s_b\left(i \frac{b + b^{-1}}{2} - \sigma \right) \ ,
\end{equation}
which has poles at $\sigma = - i b n - i b^{-1} m$ with $n,m\ge0$,
where $\sigma$ is a complexified real mass parameter. 
The Higgsing prescription will give line defects localized on two distinct $S^1$ inside the ellipsoid $S^3_b$. 

In the case of the super-conformal index, we would use 
\begin{equation}
I_{\mathrm{3d chiral}}(z,m;q) = \frac{(q z^{-1} q^{-m/2};q)_\infty}{(z q^{-m/2};q)_\infty}  \ ,
\end{equation}
which has poles at $z q^{n-m/2} =1$, where $z$ is the flavor fugacity and $m$ the background magnetic flux. 
We leave an interpretation of the Higgsing prescription for 3d indices to later work. 

The analysis of Higgsing for the index of $\CN=2$ 4d theories is analogous to the $\CN=1$ case. 
The controllable RG flows are initiated by vevs of Higgs branch operators. 
The contribution of a hypermultiplet in a standard $SU(2)_R$ representation to the index is 
\begin{equation}
I_{\mathrm{hyper}}(z;p,q,t) = \Gamma(\sqrt{t} z;p,q) \Gamma(\sqrt{t} z^{-1};p,q) \ .
\end{equation}

Constant Higgs branch vevs for operators of $SU(2)_R$ spin $k/2$ are associated to poles of the form $z = t^{k/2}$ for some flavor fugacity $z$, 
while position-dependent vevs are associated to the poles of the form $z = t^{k/2} p^r q^s$.
Thus the Higgsing prescription should be 
\begin{equation}
I_{IR}(z_a;p,q,t) = I^{-1}_{\mathrm{free}}(z_a;p,q,t) \lim_{ \prod_a z_a^{q_a} t^{k/2} p^r q^s \to 1} I^{-1}_{\mathrm{hyper}}(\prod_a z_a^{q_a} t^{(k+1)/2};p,q,t) I_{UV}(z_a;p,q,t) \ .
\end{equation}

Notice that $I_{\mathrm{free}}(z_a;p,q,t)$ should consist of full hypermultiplet contributions. 
Notice also that the residue of $I_{\mathrm{hyper}}(z;p,q)$ at $z= \sqrt{t}$ equals 
\begin{equation} I_{\mathrm{gauge}}^{-1} \Gamma(t;p,q) \equiv I_{\mathrm{vector}}^{-1}
\end{equation}
This is useful to match our formula with the results of \cite{Gaiotto:2012xa}. 

The $S^4_b$ partition function of a four-dimensional $\CN=2$ gauge theory is computed through localization~\cite{Pestun:2007rz,Hama:2012bg}
 as an integral along the imaginary part of the vectormultiplet scalar fields. The integrand includes 
 some classical contribution, a one-loop factor and the square modulus of the instanton partition function. 
 The integrand depends on the vectormultiplet scalar fields, mass parameters and gauge couplings. 
 
Crucially, the only singularities in the integrand arise from the one-loop factor: 
the instanton partition function has poles, but they cancel against the zeroes of the vectormultiplet one-loop factor. In particular, the 
poles of the partition function can be identified by the familiar combinatorics of hypermultiplet poles 
pinching the integration contour. The intstanton partition function goes along for the ride and is simply specialized to the values of Coulomb branch parameters and masses 
selected by the one-loop analysis. 

To be precise, the hypermultiplet contribution is 
\begin{equation}
Z^{S^4_b}_{\mathrm{hyper}}(m) = \frac{1}{\Upsilon(\frac{Q}{2} + m)}
\end{equation}
with the function $\Upsilon(z)$ having zeroes at $z = - b n - b^{-1} m$ and $z = b (n+1) + b^{-1} (m+1)$, 
$n,m \geq 0$ and $Q = b + b^{-1}$. 

Thus the combinatorics of poles works in parallel to the index if we identify 
\begin{equation}
\log t \to Q \qquad \qquad \log p \to b \qquad \qquad \log q \to b^{-1}
\end{equation} 
and thus the standard constant Higgs vevs due to a hypermultiplet operator 
in a spin $k/2$ representation of the R-symmetry group, located in the index at some $z=t^{\frac{k}{2}}$ 
is located in $S^4_b$ partition function at $m = \frac{k Q}{2}$. The vortex configurations 
add further multiples of $b$ and $b^{-1}$ on top of that. 

Notice that 
\begin{equation} \Upsilon(z + b) = \frac{\Gamma( b z)}{\Gamma(1- b z)} b^{1-2 b z}\Upsilon(z)
\end{equation}
should play a physical role analogous to 
\begin{equation}
I_{\mathrm{hyper}}(p z;p,q,t) = \Gamma(\sqrt{t} p z;p,q) \Gamma(\sqrt{t} z^{-1}p^{-1};p,q) = I_{\mathrm{hyper}}(z;p,q,t)\frac{\theta(\sqrt{t} z;q)}{\theta(\sqrt{t} p^{-1}z^{-1};q)}
\end{equation}
in the combinatorics of one-loop factors: the latter produces contributions which are roughly identified with the 
elliptic genus of 2d chiral multiplets, the former with $S_b^2$ partition functions 
of the same 2d chiral multiplets.  

\begin{figure}[h]
    \centering
 \includegraphics[width=.99\textwidth]{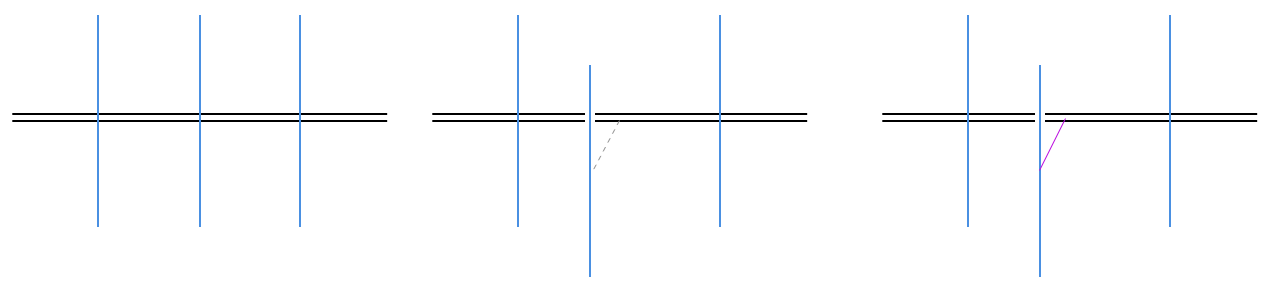}  
    \caption{The NS5-D4 brane description of a baryonic Higgs branch of an $SU(2) \times SU(2)$ quiver gauge theory which flow to a pure $SU(2)$ gauge theory in the IR. On the right, we include a D2 brane segment corresponding to a position-dependent Higgs branch}
    \label{fig:zeroa}
\end{figure}

In the context of class S theories, an important class of Higgs branch deformations
is associated to poles in the flavor fugacities for flavor symmetries associated with a single puncture of the 
UV Riemann surface defining the theory. These Higgs branch deformations may erase a puncture, or interpolate between different types of punctures
\cite{Gaiotto:2012uq}. The corresponding position-dependent Higgs branch deformations produce an interesting class of surface defects,
which are also associated to punctures on the UV Riemann surface, and expected to correspond to co-dimension four defects in the six-dimensional UV
SCFT. From the point of view of linear quiver gauge theories, these are either simple baryonic Higgs branch deformations 
which turn on a vev for a bi-fundamental field, thus Higgsing two consecutive gauge nodes to the diagonal subgroup and removing a simple puncture on the UV curve, 
or mesonic Higgs branch deformations involving the fundamental fields at one end of the quiver. See Figure \ref{fig:zeroa}.

Using the relation between the Higgsing prescription for the index and the $S^4_b$ partition function
we recover a well-known aspect of the AGT correspondence: the specialization of mass parameters associated to the Higgsing 
of punctures in $S^4_b$ coincides with the specialization of Liouville/Toda momenta which relates generic, semi-degenerate or degenerate 
punctures among themselves.

\begin{figure}[h]
    \centering
    \includegraphics[width=.99\textwidth]{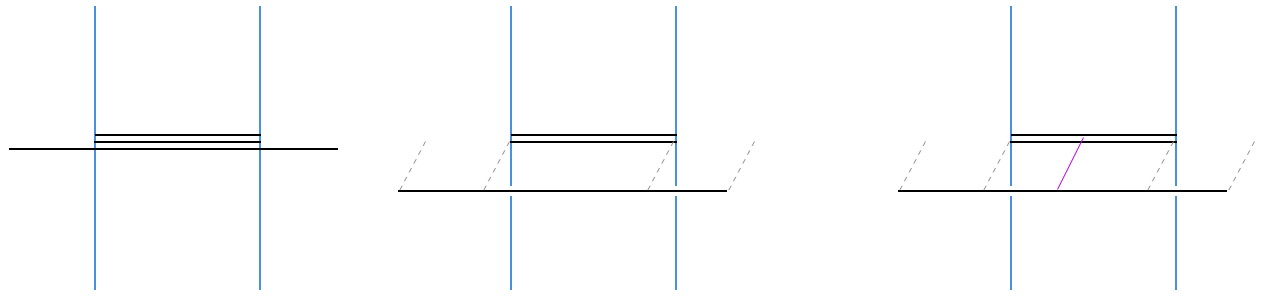} 
    \caption{The NS5-D4 brane description of a baryonic Higgs branch of an $SU(3)$ theory with two flavors which flow to a pure $SU(2)$ gauge theory in the IR. On the right, we include a D2 brane segment corresponding to a position-dependent Higgs branch}
    \label{fig:zerob}
\end{figure}

These deformations by no means exhaust the possible Higgs branch deformations of quiver gauge theories or class S theories. 
At the opposite end of the spectrum, one can consider Higgs branch deformations which 
affect the whole Riemann surface in a class S theory, flowing from a $A_k$- to an $A_{k-1}$-type theory with the ``same'' Riemann surface. 
In the language of brane constructions, the mesonic Higgs branch vevs correspond to separating one (or more) 
M5 branes from the stack of $k+1$ M5 branes which engineers the four-dimensional theory. This type of Higgs branch is important, for example, 
in the work \cite{Aganagic:2014oia}. From the point of view of linear quivers, these are deformations which give vevs to composite operators 
which stretch across the whole quiver. These Higgs branch deformations are rather unexpected in the context of AGT:
they reduce, say, conformal blocks and correlation functions of some $A_{k}$ Toda theory to the ones for a 
$A_{k-1}$ Toda theory. See Figure \ref{fig:zerob}.

\begin{figure}[h]
    \centering
    \includegraphics[width=0.32\textwidth]{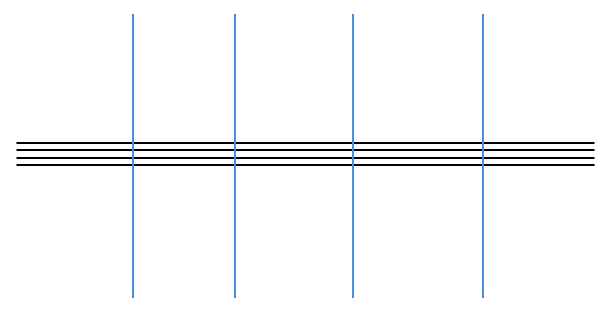}  \includegraphics[width=0.32\textwidth]{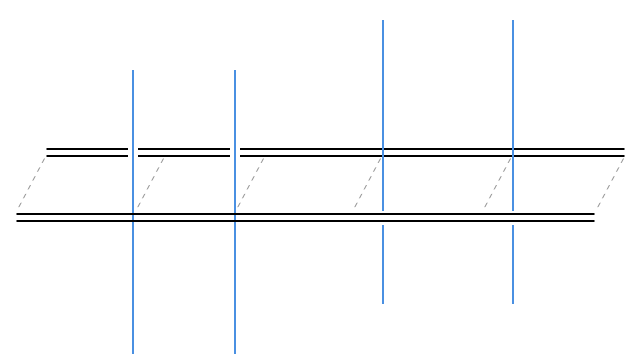}  \includegraphics[width=0.32\textwidth]{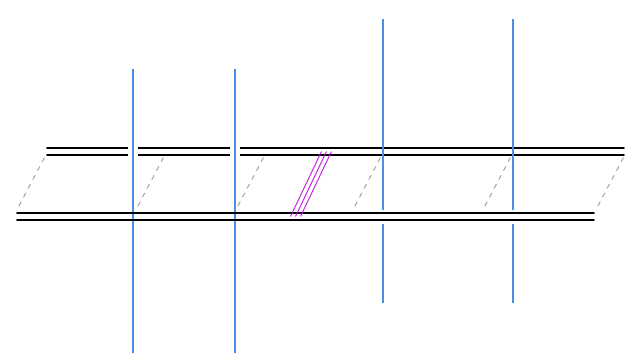}
    \caption{The NS5-D4 brane description of a particular Higgs branch of a $SU(4)^3$ linear quiver gauge theory. On the left: the un-Higgsed theory. In the middle: a Higgs branch deformation which flows to the product of two $SU(2)$ SQCD theories, each with four flavors. On the right: a position dependent Higgs branch vev flows to a surface defect associated to D2 branes stretched between the two sub-systems.}
    \label{fig:one}
\end{figure}

There are many other possible Higgs branches for class S theories, which correspond to separating the stack of M5 branes into two sub-stacks,
while distributing among the two stacks the transverse M5 branes which define the punctures. The result in the IR are two decoupled class S theories. 
We give a graphic depiction of such general Higgs branch in Figure \ref{fig:one}.
Position dependent versions of these Higgs branches will create surface defects which couple to the two IR sub-theories. 

Many four-dimensional $\CN=2$ gauge theories admit a lift to a five-dimensional gauge theory, with a 5d SCFT UV completion. 
The lift is straightforward for quiver gauge theories, but is also available for many non-Lagrangian examples, including 
general class S trinion theories and possibly any genus 0 class S theory \cite{Benini:2009gi,Aganagic:2014oia}. 
The $S^4_b$ partition function can be recovered as a limit of the 5d superconformal index
(though this is very challenging for non-Lagrangian examples). For quiver gauge theories, 
all our calculations can be done with the same ease in four or five dimensions, 
and some aspects are more intuitive in five dimensions. We will thus 
focus on five-dimensional theories. 

The supersymmetric index of a five-dimensional hypermultiplet is 
\begin{equation}
I_{\mathrm{hyper}}(z;p,q) = \frac{1}{(\sqrt{p q} z;p,q)_\infty (\sqrt{p q} z^{-1};p,q)_\infty} \ ,
\end{equation}
where $z$ is the fugacity for the $U(1)$ flavor symmetry (the Cartan sub-algebra of the actual $Sp(1)$ flavor symmetry of a free hypermultiplet)
and $p,q$ are fugacities for linear combinations of rotations and R-symmetry charges. 
This is the obvious ``lift'' of $Z^{S^4_b}_{\mathrm{hyper}}$, and has a similar structure of zeroes and poles, 
with linear combinations of $b$ and $b^{-1}$ being replaced by monomials in $p$ and $q$. 

The twisted $S^1 \times S^4$ partition function of a five-dimensional supersymmetric gauge theory,
which often can be interpreted as the index of some 5d SCFT which is a UV completion of the gauge theory, 
has the structure of a contour integral over the Cartan torus of the one-loop factor for hypermultiplets and gauge multiplets,
multiplying the ``square modulus'' of the instanton partition function. 

The standard definition of the conjugation operation on the instanton partition function involves inverting all the gauge and flavor fugacities, including the 
fugacity associated to the instanton charge, which functions as an instanton-counting parameter in the 
instanton partition function. The instanton partition function is defined as a power series in the 
instanton-counting parameter, but in order to make sense of the index and of the conjugation operation, 
one needs to re-expand it as a power series in positive powers of the fugacities $p$ and $q$. 
The coefficients of individual powers of $p$ and $q$ are rational functions in the instanton counting parameter, and thus the 
standard conjugation prescription gives meaningful answers. 

The intuition behind the standard definition of conjugation is that the flavor and gauge fugacities are pure phases, while $p$ and $q$ 
are real and smaller than $1$. Once we start considering 3d/5d systems and Higgsing, we will often find ourselves shifting
flavor and gauge fugacities by powers of $p$ and $q$. These operations do not ``commute'' with the standard notion of conjugation 
and thus would require one to give separate prescriptions for fugacities which appear in the instanton and anti-instanton 
contributions to the index of 3d/5d systems. This is cumbersome and sometimes problematic. 

We can avoid that if we re-define slightly the notion of complex conjugation, by inverting $p$ and $q$ as well
{\it before} we expand the anti-instanton contribution into positive powers of $p$ and $q$. This operation is meaningful as the contributions 
of fixed instanton number to the instanton partition function are rational functions in $p$ and $q$. 
This operation actually does nothing to the standard instanton partition function, so our modified definition does not change the usual answer. 
This fact related to the non-Abelian nature of the R-symmetry and rotation symmetry groups in $5d$. 

The modified definition of complex conjugation commutes with shifts of flavor and gauge fugacities by powers of $p$ and $q$ 
and is thus rather convenient for $3d/5d$ and Higgsing calculations. For 3d/5d systems the unbroken R-symmetry is Abelian, and 
the inversion of $p$ and $q$ will act non-trivially. 

Even the one-loop contributions to the instanton partition function can be often written down
(in multiple ways) as a ``square modulus'' of some simpler expression, 
to be identified with the one-loop contribution to the instanton partition function. 
The inversion of $p$ and $q$ is a bit formal, but can be done readily at the level of the plethystic logarithm. 

For example, the hypermultiplet index can be written as a square modulus: 
\begin{equation}
I_{\mathrm{hyper}}(z;p,q) = \frac{1}{|(\sqrt{p q} z;p,q)_\infty|^2} \ .
\end{equation}
Notice that the plethystic logarithm of $(\sqrt{p q} z;p,q)_\infty^{-1}$, i.e. $\frac{\sqrt{p q} z}{(1-p)(1-q)}$
is invariant under $p \to p^{-1}$ and $q \to q^{-1}$, as expected. 

On the other hand, the index of a free 3d chiral multiplet written as a square modulus
\begin{equation}
I_{\mathrm{3d chiral}}(z;q) = \frac{(q z^{-1};q)_\infty}{(z;q)_\infty} = \frac{1}{|(z;q)_\infty|^2}  
\end{equation}
shows a non-trivial application of the conjugation rule: 
the plethystic logarithm of $(z;q)_\infty^{-1}$ is $\frac{z}{1-q}$ which is sent by inversion of all fugacities to 
$-\frac{q z^{-1}}{1-q}$.

The poles arising from a standard constant Higgs vevs due to a hypermultiplet operator 
in a spin $k/2$ representation of the R-symmetry group, located in the 4d index at some $z=t^{\frac{k}{2}}$ 
and in the 4d $S^4_b$ partition function at $m = \frac{k Q}{2}$, will be located in the 5d index at $z = p^{k/2} q^{k/2}$.
Vortices will add further powers of $p$ and $q$ on top of that. 

Thus the Higgsing prescription should be 
\begin{equation}
I_{IR}(z_a;p,q) = I^{-1}_{\mathrm{free}}(z_a;p,q) \lim_{ \prod_a z_a^{q_a} p^{r+k/2} q^{s+k/2} \to 1} I^{-1}_{\mathrm{hyper}}(\prod_a z_a^{q_a} p^{(k+1)/2} q^{(k+1)/2};p,q,t) I_{UV}(z_a;p,q) \ .
\end{equation}
Notice that we have the usual relation between shifted indices for hypermultiplets and 
indices for chiral multiplets in two fewer dimensions: 
\begin{equation}
I_{\mathrm{hyper}}(p z;p,q) = I_{\mathrm{hyper}}(z;p,q) I_{\mathrm{3d chiral}}(\sqrt{p q} p^{-1} z^{-1};q) \ .
\end{equation}
Thus the one-loop combinatorics upon Higgsing will produce precisely the 3d analogues of the 
2d chiral multiplet partition functions and indices we encountered when Higgsing 4d gauge theories. 

\subsection{Examples of Higgsing in four-dimensional  $\CN=1$ gauge theories}
In this section we consider a simple example, consider $SU(N_c)$ SQCD with $N_f$ flavors. 
We are not aiming to a full discussion of the physical properties to the relevant $(0,2)$ half-BPS defects, 
as it would require a careful analysis beyond the scope of this paper. 
We only aim to show some basic features of the Higgsing procedure on the superconformal index which persist 
in theories with more supersymmetry and higher dimension (and presumably lower dimension as well, 
though we will not explore that direction).  

Including the vectormultiplet contributions, we get the index of  SQCD 
\begin{equation}
I^{\CN=1}_{N_c,N_f} = \frac{(p;p)^{N_c-1}_\infty (q;q)^{N_c-1}_\infty}{N_c!} \oint \prod_{j=1}^{N_c-1} \frac{dz_i}{2 \pi i z_j} \frac{\prod_{i,a}\Gamma(s_a^{-1} z_i;p,q)\Gamma(t_a z_i^{-1};p,q)}{\prod_{i\neq j} \Gamma(z_i z_j^{-1};p,q)}
\end{equation}
with $\prod_i z_i =1$ and $\prod_a t_a = (pq)^{N_f-N_c} \prod_a s_a$ to set to zero the anomalous axial symmetry. 

We can consider two types of poles for the index: mesonic and baryonic poles. If we pick a single meson operator, 
say $\tilde Q^{N_f} Q_{N_f}$, and give it a vev, we expect to Higgs the theory down to SQCD with $N_f$ and $N_c$ lower by one unit. 
This operator has fugacity $t_{N_f} s^{-1}_{N_f}$. The corresponding pole arises from pinching the integration contour 
in the integral expression for the index: if we adjust $t_{N_f} = s_{N_f}$, any of the pair of poles of the integrand 
at $s_{N_f}  =z_i$ and $t_{N_f} = z_i$ pinch the contour. 

The choice of which $z_i$ appears in the pole is immaterial, it only gives an overall factor of $N_c$, which reduces the $N_c!$ in the denominator to the expected $(N_c-1)!$. Thus the residue is easily computed 
\begin{align}
\hspace{-1cm}\lim_{t_{N_f} \to s_{N_f}} &I_{\mathrm{chiral}}(t_{N_f} s_{N_f}^{-1};p,q) I^{\CN=1}_{N_c,N_f}(s_a ,t_a ;p,q) = \cr &\left[\prod_{a}\Gamma(s_a^{-1} t_{N_f};p,q)\Gamma(t_a s_{N_f}^{-1};p,q)\right] I^{\CN=1}_{N_c-1,N_f-1}(s_a s_{N_f}^{-\frac{1}{N_c-1}},t_a s_{N_f}^{-\frac{1}{N_c-1}};p,q) \ .
\end{align}
We see the expected free chirals of fugacity $s_a^{-1} s_{N_f}$ and $t_a t_{N_f}^{-1}$ which arise as Goldstone bosons for the breaking of $SU(N_f)^2$ to $SU(N_f-1)^2$. In order to obtain the index of SQCD with $N_f-1$ flavors and $N_c-1$ colours we 
need to subtract them off. 

For a general pole at $t_{N_f} = p^r q^s s_{N_f}$ we find $(r+1)(s+1)$ contributions, at $z_{N_c} = p^{r'} q^{s'} s_{N_f}$
with $r'=0, \cdots, r$, $s'=0, \cdots, s$. We can specialize to $s=0$ to study the index of surface defects, 
rather than the index for the intersection  of two orthogonal defects. Thus we have a sum of $r+1$ terms, 
labelled by $r'=0, \cdots, r$. 

The interpretation of these terms is rather obvious: they are equivariant fixed points in the 
moduli space of ``vortex'' configurations, where $Q_{N_f}$ has a zero of order $r'$ and $\tilde Q^{N_f}$ 
has a zero of order $r-r'$. The answer thus appears as a calculation of an equivariant elliptic genus 
of the vortex degrees of freedom as a sum over fixed points. For each choice of $r'$, the imperfect cancellations between 
$\Gamma$ functions at numerator and denominator of the integrand will 
give rise to ratios of products of $\theta$ functions. These $\theta$ functions can be interpreted as the 
contributions of 2d $(0,2)$ chiral or Fermi multiplets associated to the tangent space at the equivariant fixed point. 

The precise answer depends a bit on the choice of how to subtract the 
Goldstone boson contributions. We get the nicest formula if we subtract  $\prod_{a}\Gamma(s_a^{-1} t_{N_f};p,q)\Gamma(t_a s_{N_f}^{-1};p,q)$,
whose arguments are the fugacities of the mesonic operators $\tilde Q^{a} Q_{N_f}$ and $\tilde Q^{N_f} Q_{a}$. Alternative choices 
involving arguments shifted by powers of $p$ differ by $\theta$ functions which can be interpreted as the contribution of extra 2d $(0,2)$ chiral or Fermi
multiplets which carry no gauge charges, possibly coupled to the defect degrees of freedom by some fermionic superpotential couplings.
Finding the correct choice goes beyond the scope of this paper.

The part of the answer which depends on gauge fugacities involves the product of $\theta$ functions
\begin{equation}
\prod_{k=1}^{r'}\theta(p^{-k} s_{N_f}^{-1} z_i;q) \prod_{k=r'}^{r-1}\theta(p^{k} s_{N_f} z_i^{-1};q) \ ,
\end{equation}
which can be interpreted naively as $r'$ fundamental Fermi multiplets and $r-r'$ anti-fundamental Fermi multiplets,
with various charges under the generator of rotations around the defect.

The part of the answer which depends on flavour gives
\begin{equation}
\prod_{k=1}^{r'}\theta(p^{-k} t_a  s_{N_f}^{-1};q)^{-1} \prod_{k=r'}^{r-1}\theta(p^{k} s_a^{-1}s_{N_f};q)^{-1} \ ,
\end{equation}
which can be interpreted naively as $r'$ 2d chiral multiplets in the fundamental of one $SU(N_f-1)$ and and $r-r'$ 2d chiral multiplets
the anti-fundamental of the other $SU(N_f-1)$, with various charges under the generator of rotations around the defect.
Furthermore, we have a final collection of theta functions whose arguments are powers of $p$ only. 
We only give here the plethystic logarithm of these terms: $1-\frac{(1-p^{1-r} q)(1-p^{r'})(1-p^{r-r'})}{(1-p)(1-q)}$

The fugacities of the 2d multiplets seems 
correct for $E$-type coupling involving the Fermi multiplets and the product of the chiral multiplets and 
(anti-)quarks restricted to the defect. 

For example, for $r=1$ we have two contributions to the residue. At $z_{N_c} = s_{N_f}$ we get: 
\begin{align}
 \prod_{a}\theta(s_a^{-1} s_{N_f};q)^{-1} \frac{(p;p)^{N_c-2}_\infty (q;q)^{N_c-2}_\infty}{(N_c-1)!} \oint \prod_{j=1}^{N_c-2} \frac{dz_i}{2 \pi i z_j} \mathrm{Int}(z_i,s_a,t_a;p,q)\prod_{i} \theta(s_{N_f} z_i^{-1};q)
\end{align}
with $\mathrm{Int}(z_i,s_a,t_a;p,q)$ 
being the standard integrand for SQCD with $N_c-1$ colours and $N_f-1$ flavors. 
At $z_{N_c} = p s_{N_f}$ we get: 
\begin{align}
\prod_{a}  \theta(p^{-1} t_a s_{N_f}^{-1};q)^{-1} \frac{(p;p)^{N_c-2}_\infty (q;q)^{N_c-2}_\infty}{(N_c-1)!} \oint \prod_{j=1}^{N_c-2} \frac{dz_i}{2 \pi i z_j} \mathrm{Int}(z_i,s_a,t_a;p,q)\prod_{i} \theta(z_i p^{-1} s_{N_f}^{-1};q) \ .
\end{align}

Notice that the contour integrals over the remaining $z_i$ are done with the constraint 
$\prod_i z_i =p^{-r'} s_{N_f}^{-1}$ and $p^r \prod_a t_a = (pq)^{N_f-N_c} \prod_a s_a$.
In order to get to a standard $SU(N_c-1)$ contour integral one should shift 
$z_i \to p^{-r'/N_c} s_{N_f}^{-1/N_c} z_i$. This shift can be reabsorbed into a simultaneous shift of 
$s_a$, $t_a$, so that it only affects the 2d degrees of freedom.

We can re-do the calculation in the magnetic dual frame. The meaning of the mesonic Higgs branch in that frame
is well understood \cite{Seiberg:1994pq}: we are giving a vev to an elementary 
chiral operator which enters linearly in the superpotential, coupled to the magnetic quarks and anti-quarks. 
Thus it gives a mass to a quark-anti-quark pair, and induces a flow from $SU(N_f-N_c)$ with $N_f$ flavors
to $SU(N_f-N_c)$ with $N_f-1$.

The magnetic expression for the index can be written as 
\begin{align}
I^{\CN=1}_{N_c,N_f} = \prod_{a,b} \Gamma(t_a s_b^{-1};p,q)& \frac{(p;p)^{N_f-N_c-1}_\infty (q;q)^{N_f-N_c-1}_\infty}{(N_f-N_c)!} \cdot
\cr & \oint \prod_{j=1}^{N_f-N_c-1} \frac{d\tilde z_i}{2 \pi i \tilde z_j} \frac{\prod_{i,a}\Gamma(s_a \tilde z_i;p,q)\Gamma(p q t^{-1}_a \tilde z_i^{-1};p,q)}{\prod_{i\neq j} \Gamma(\tilde z_i \tilde z_j^{-1};p,q)}
\end{align}
with $\prod_i \tilde z_i \prod_a s_a =1$. 
We see the cancellation of the quark and anti-quark contributions in the integrand if we set $t_{N_f} = s_{N_f}$. 

A position-dependent vev has a similar effect, but the quark-anti-quark pair does not cancel out completely. 
If $s=0$ we are left with 
\begin{equation}
\prod_{k=0}^{r-1} \theta(p^k s_{N_f} \tilde z_i;q)^{-1} \ ,
\end{equation}
which can be interpreted as a coupling of the magnetic dual gauge fields to fundamental chiral multiplets. 
Here there is no sum over fixed points. 

Thus we get a neat duality between surface defects in the electric and magnetic Seiberg dual theories, 
up to the physics subtleties sketched above. It would be interesting to check if the duality still holds when replacing 
the $p^k s_{N_f} \tilde z_i$ fugacities above by a more general collection of $r$ $2d$ fugacities $s_{a} \tilde z_i$. Presumably this can be done through 
a sequence of $r=1$ mesonic Higgsings starting from SQCD with a large number of colours and flavors. 

The second possibility is to look at RG flows initiated by a baryonic operator vev. 
Essentially, we take an $N_c \times N_c$ subset of quarks and give it a vev proportional to the 
identity. This Higgses completely the gauge group and breaks the flavor group to $SU(N_f) \times SU(N_c)\times SU(N_f-N_c)$. 
The corresponding manipulations of the index 
with or without vortices may be found in \cite{Chen:2014rca}. 

The appropriate pole arises from the combination of poles in the integrand of the form 
$z_i  = s_i p^{r_i} q^{h_i}$, which give a pole for the index at the value of the baryon fugacity 
$\prod_i s_i^{-1} = p^r q^s$ with $r=\sum r_i$ and $s=\sum_i h_i$. 

For $r=0$,$s=0$ we are left with 4d chirals in bifundamental representations of $SU(N_f) \times SU(N_c)$ and $SU(N_c)\times SU(N_f-N_c)$.
For $s=0$ and general $r$ we get a sum of many terms, each including extra 2d multiplets. It seems feasible to reproduce this sum from the elliptic genus of certain 2d gauge theories, which engineer the vortex moduli spaces of SQCD, as in \cite{Chen:2014rca}. 
Similar considerations likely apply to the meson Higgs branch vortices. We leave the analysis to future work.

\subsection{Examples of Higgsing in four-dimensional $\CN=2$ gauge theories}
If we look at the index of $\CN=2$ SQCD we can find both mesonic and baryonic poles. 
In order to have a super-conformal field theory, we need $N_f = 2 N_c$. We will not set 
$N_f = 2 N_c$ explicitly in the formulae below, as we are mostly interested in 
comparing the index formulae with $S^4_b$ partition function formulae, which do not have such a restriction. 

The mesonic branch vev involves two hypermultiplet flavors at the time, as the F-term relations 
require the meson vev to be a nilpotent matrix. In $\CN=1$ language, we are turning on 
a vev for $M^{N_f}_{N_f-1} = \tilde Q_{N_f-1} Q^{N_f}$. In the $\CN=2$ index,
\begin{equation}
I^{\CN=2}_{N_c,N_f} = \frac{(p;p)^{N_c-1}_\infty (q;q)^{N_c-1}_\infty}{(N_c!)\Gamma(t ;p,q)^{N_c-1}} \oint \prod_{j=1}^{N_c-1} \frac{dz_i}{2 \pi i z_j} \frac{\prod_{i,a}\Gamma(\sqrt{t} s_a^{-1} z_i;p,q)\Gamma(\sqrt{t} s_a z_i^{-1};p,q)}{\prod_{i\neq j} \Gamma(z_i z_j^{-1};p,q)\Gamma(t z_i z_j^{-1};p,q)}
\end{equation}
we set $s_{N_f-1} = t s_{N_f}$ and pinch the contour at $z_{i} = \sqrt{t}s_{N_f}$, i.e. $s_{N_f-1} = \sqrt{t}z_{i}$.
After the usual simplifications, we are left with the index of $N_c-1$ SQCD with $N_f-2$ flavors,
up to the $\CN=2$ Goldstone boson contributions $\prod_{a}\Gamma(t s_{N_f} s_a^{-1};p,q)\Gamma(s_{N_f}^{-1} s_a;p,q)$.

In order to study a position-dependent Higgs branch vev, we pick poles at $s_{N_f-1} = t p^r s_{N_f}$ and $z_{N_c} = \sqrt{t}p^{r'} s_{N_f}$.
We subtract the standard normalization factor as for the definition above and the integrand for SQCD with $N_c-1$ and $N_f-2$ flavors.
The part of the answer which depends on the gauge fugacities is a collection of $\theta$ functions which can be interpreted 
as a collection of fundamental and anti-fundamental $(2,2)$ chiral multiplets from the tangent space to the equivariant fixed point in the vortex moduli space:
\begin{equation}
\prod_i \prod_{k=0}^{r'-1}\frac{\theta(\sqrt{t}p^k s_{N_f} z_i^{-1};q)}{ \theta(\sqrt{t}p^{-k-1} s_{N_f}^{-1} z_i;q)}   \prod_{k=r'}^{r-1}\frac{\theta(1/\sqrt{t}p^{-k-1} s_{N_f}^{-1} z_i;q)}{\theta(t\sqrt{t}p^{k} s_{N_f} z_i^{-1};q)} \ .
\end{equation}

We have two choices of how to subtract the Goldstone bosons: we can either subtract 
$\prod_{a}\Gamma(t s_{N_f} s_a^{-1};p,q)\Gamma(s_{N_f}^{-1} s_a;p,q)$ or $\prod_{a}\Gamma(t p^r s_{N_f} s_a^{-1};p,q)\Gamma(p^{-r} s_{N_f}^{-1} s_a;p,q)$. 
More symmetric choices would not be compatible with $\CN=2$ 
supersymmetry, unless we break some extra flavor symmetry. 

If we go ahead with the latter, we get a contribution depending on flavor fugacities only: 
\begin{equation}
\prod_a \prod_{k=r'}^{r-1}\frac{\theta(t p^k s_a^{-1}  s_{N_f};q)}{\theta(p^{-k-1}s_a s_{N_f}^{-1};q)}
\end{equation}
compatible with superpotential couplings involving the $(2,2)$ chirals with gauge charges and the hypermultiplets restricted to the defect. 
Finally, we get some extra $(2,2)$ chirals with no flavor fugacities, corresponding to the plethystic exponent
\begin{align}
1-\frac{(\sqrt{t}- p q/\sqrt{t})(1-p^{r'})(1-p^{r-r'})(\sqrt{t} +p^{-r}/\sqrt{t})}{(1-p)(1-q)} \ .
\end{align}

Notice that the remaining gauge fugacities satisfy $\sqrt{t}p^{r'} s_{N_f} \prod_i z_i =1$, 
and thus we need to shift them to bring them back to $SU(N_c-1)$ fugacities. 
The shift can be reabsorbed into a shift of the $s_a$ and thus only affects the charges of 2d degrees of freedom. 

If we specialize to $r=1$, we get a sum of two terms. 
The $r'=0$ term involves anti-fundamental chirals accompanied by extra flavoured chirals in a fundamental representation of the flavor group:
\begin{equation}
\prod_i \frac{\theta(1/\sqrt{t}p^{-1} s_{N_f}^{-1} z_i;q)}{\theta(t\sqrt{t} s_{N_f} z_i^{-1};q)} \prod_a \frac{\theta(t s_a^{-1}  s_{N_f};q)}{\theta(p^{-1}s_a s_{N_f}^{-1};q)}
\end{equation}
with $\sqrt{t}p^{r'} s_{N_f} \prod_i z_i =1$.
The $r'=1$ term involves fundamental chirals only: 
\begin{equation}
\prod_i \frac{\theta(\sqrt{t} s_{N_f} z_i^{-1};q)}{ \theta(\sqrt{t}p^{-1} s_{N_f}^{-1} z_i;q)}  \ .
\end{equation}

The calculations above have a very simple interpretation in terms of brane systems. 
The SQCD theory can be engineered by a system of two NS5 branes, with $N_c$ D4 segments 
stretched between them, and $N_f$ semi-infinite D4 branes ending on the system either from the left of from the right \cite{Witten:1997sc}.

The mesonic Higgs branch is most easily understood by having a semi-infinite D4 brane on the left, and one on the right. 
Joining them with a D4 segment into a single D4 brane which can be separated from the system 
corresponds to the mesonic Higgs branch described above. 
To keep the residual $U(N_f-2)$ flavor symmetry unbroken, we keep all the remaining semi-infinite D4 brane segments 
on the right of the system. 

The background vortices which engineer the position-dependent Higgs vevs become extra D2 brane 
segments in the IIA setup, stretched between the D4 brane and the rest of the system, as in Figure \ref{fig:zerob}. 
The D2 branes support a three-dimensional $U(r)$ gauge theory, but as they end on a single D4 brane 
the three-dimensional gauge theory degrees of freedom are subject to a generalized Dirichlet boundary condition, with a full Nahm pole. 

The fixed contributions to the index have an obvious interpretation as arising from configurations where 
$r-r'$ D2 branes lie close to the first NS5 brane and $r'$ lie close to the second NS5 brane: 
the $(2,2)$ chirals with gauge or flavor charges arise as $2-4$ strings stretched across the NS5 brane. 

The powers of $p$ in their fugacities can likely be understood as due to the Nahm pole at the other end of the 
D2 brane segment, which is invariant under a combination of rotations in the plane orthogonal to the D2 brane and parallel to the D4 brane, and gauge transformations in the D2 brane world-volume theory 
generated by the $t^3$ generator of the $\mathfrak{su(2)}$ embedding associated to the Nahm pole. 

Finally, the collection of extra $(2,2)$ chirals with no gauge nor flavor charges may presumably be understood 
as accounting for the degrees of freedom arising upon Higgsing the D2 brane world volume theory to 
$U(r') \times U(r-r')$ in the presence of the Nahm pole. 

In comparing the field theory analysis with the brane setup, it is important to remember that the 
four-dimensional degrees of freedom at the brane intersections are coupled to the
five-dimensional gauge theories supported on the bundles of semi-infinite D4 branes. 
The five-dimensional gauge fields are IR free but still impose constraints such as the F- and D-term constraints 
on the moment maps for the four-dimensional degrees of freedom. As a result, while the Higgs branch 
associated to a pair of flavors engineered semi-infinite branes on opposite sides of the system is visible as a set of D4 brane segments reconnecting and separating from the NS5 branes, the Higgs branch associated to
a pair of flavors engineered by semi-infinite branes on the same side of the system is not. 

The Higgs branch becomes visible if we de-couple the five-dimensional degrees of freedom, 
say by imposing Dirichlet boundary conditions on them, i.e. terminating each D4 brane on a 
separate transverse D6 brane. Mesonic Higgsing removes a D4 brane segments stretched 
between a pair of D6 branes and replaces the Dirichlet b.c. for the D4 branes world volume theory 
with a minimal Nahm pole. Background vortices must correspond to some instanton configurations on the 
D6 branes world volume which carries D2 brane charge, or equivalently some 
co-dimension two defect living at a Nahm pole boundary of five-dimensional gauge theory \cite{Witten:2011zz}

\subsection{Higgsing in five-dimensional $\CN=1$ gauge theories}

As an example, the index of five-dimensional SQCD takes the form 
\begin{align}
&I_{\mathrm{SQCD}}^{N_c,N_f}(\lambda,\mu_a;p,q) =  \frac{(p;p,q)_\infty^{N_c-1} (q ;p,q)_\infty^{N_c-1}}{N_c!} \cdot
\cr &\oint \prod_i \frac{dz_i}{2 \pi i z_i} \frac{\prod_{i \neq j}(z_i z_j^{-1};p,q)_\infty (p q z_i z_j^{-1};p,q)_\infty}{\prod_{i,a} (\sqrt{p q} z_i \mu_a^{-1};p,q)_\infty (\sqrt{p q} z_i^{-1} \mu_a;p,q)_\infty } |Z_{\mathrm{inst}}(z_i, \mu_a,\lambda;p,q)|^2\end{align}
with $\prod_i z_i=1$. 

The usual mesonic branch poles are located at $\mu_{N_f-1} = p^{r+1} q^{s+1} \mu_{N_f}$, 
and arise from pinching contours at $z_i = \sqrt{p q} p^{r'} q^{s'} \mu_{N_f}$.
We will focus on the poles at $\mu_{N_f-1} = p^{r+1} q \mu_{N_f}$, adding up contributions from $z_{N_c} = \sqrt{p q} p^{r'} \mu_{N_f}$.
Again, we interpret these contributions as equivariant fixed points in some moduli space of vortex configurations. 

At first, we can look at the one-loop factors. Besides the one-loop factor for the IR SQCD with 
$N_c-1$ colours and $N_f-2$ flavors, we find a collection of 3d chiral indices. 
The part which depends on gauge fugacities can be rearranged to the familiar 
contribution of 3d chirals in fundamental and anti-fundamental representations: 
\begin{align}
\prod_i \prod_{k=r'+1}^r I_{\mathrm{3d chiral}}(p^{-k}/\sqrt{p q} z_i \mu_{N_f}^{-1};q )  \prod_{k=0}^{r'-1} I_{\mathrm{3d chiral}}(p^k \sqrt{p q}z_i^{-1} \mu_{N_f}) \ .
\end{align}

After subtracting the Goldstone boson contribution $I_{\mathrm{hyper}}(\mu_a \mu_{N_f-1}^{-1}\sqrt{p q})$
(there is as before an ambiguity between subtracting that or $I_{\mathrm{hyper}}(\mu_a \mu_{N_f}^{-1}/\sqrt{p q})$)
the part which depends on the flavor fugacities only gives 
the contribution of extra 3d chirals 
\begin{align}
\prod_{a} \prod_{k=r'+1}^r I_{\mathrm{3d chiral}}( p^k q \mu_{N_f} \mu_a^{-1}) \ .
\end{align}

Finally, we get other 3d chiral contributions with no gauge or flavor fugacities from the remaining plethystic logarithm:
\begin{align}
1- \frac{(p^{-r} +p q)(1-p^{r'})(1-p^{r-r'})}{(1-p)(1-q)} \ .
\end{align}

For $r=0$, we have reduced the one-loop factors to the ones for the IR SQCD with 
$N_c-1$ colours and $N_f-2$ flavors. In order to reproduce the full index for that theory, 
we should also verify that the instanton contributions match when we specialize the gauge and flavor fugacities appropriately.  

For simplicity, we can work with an $U(N_c)$ instanton partition function.
The partition function at instanton number $k$ is written as an ADHM contour integral.
which contains the combination 
\begin{equation}
\frac{\rho_I^{N_c-N_f/2}\prod_a \mu_a^{-1/2}( \rho_I- \mu_a)}{\prod_i(\sqrt{p q}^{-1}\rho_I- z_i)(\sqrt{p q} \rho_I- z_i)} \ ,
\end{equation}
where $\rho_I$ ate the integration variables for the ADHM contour integral.

Once we specialize to $\mu_{N_f-1} = p q \mu_{N_f}$ and $z_{N_c} = \sqrt{p q} \mu_{N_f}$
the corresponding factors in the expression above
\begin{equation}
\frac{(\rho_I- \mu_{N_f})( \rho_I- p q \mu_{N_f})}{(\sqrt{p q}^{-1}\rho_I- \sqrt{p q} \mu_{N_f})(\sqrt{p q} \rho_I- \sqrt{p q} \mu_{N_f})}
\end{equation}
cancel out.

On the other hand, if we specialize to $\mu_{N_f-1} = p^{r+1} q \mu_{N_f}$ and $z_{N_c} = \sqrt{p q} p^{r'} \mu_{N_f}$
we get 
\begin{equation}
\frac{( \rho_I- \mu_{N_f})( \rho_I- p^{r+1}q \mu_{N_f})}{(\rho_I- p^{r'+1}q\mu_{N_f})( \rho_I- p^{r'}\mu_{N_f})} \ .
\end{equation}
We can re-write that as a telescoping product 
\begin{align}
\prod_{k=r'+1}^r \frac{( \rho_I- p^{k+1}q \mu_{N_f})}{(\rho_I- p^{k}q\mu_{N_f})} 
\prod_{k=0}^{r'-1} \frac{(\rho_I- p^k\mu_{N_f})}{(\rho_I- p^{k+1}\mu_{N_f})} 
\end{align}
and tentatively identify each factor as the contribution to the 
equivariant index of 3d chiral matter in the fundamental 
\begin{equation}
\frac{( \rho_I- \mu)}{(\rho_I- \mu p^{-1})} 
\end{equation}
and anti-fundamental representations respectively:
\begin{equation}
\frac{(p q \rho_I- \mu )}{(p q \rho_I- p \mu)}  \ .
\end{equation}

As in the four-dimensional case, we can tentatively understand the different contributions to the index 
by engineering the five-dimensional gauge theory as a set of D5 branes intersecting two NS5 branes. 
Higgsing can be described by separating one full D5 brane from the system, Higgsing with vortices 
by stretching $r$ D3 brane segments between the D5 brane and the rest of the system. 
The same considerations apply concerning the effect of generalized Dirichlet boundary conditions 
on the D3 brane world-volume theory. 

Let us briefly comment on the structure of the symmetry and the superconformal index in the 3d/5d coupled system.
The Higgsing with defects breaks the $F(4)$ superconformal symmetry of the five-dimensional CFT to the $\mathcal{N}=2$ superconformal symmetry of the 3d theory on the defect. The bosonic subalgebra $SO(2,5)\times SU(2)_R$ is broken to $SO(2,3)\times U(1)_R\times U(1)_f$ subalgebra where $SO(2,3)$ is the conformal algebra of the 3d theory. $U(1)_R$ and $U(1)_f$ are the diagonal and off-diagonal combinations of $SO(2)$ and $SU(2)_R$ symmetries, respectively, where $SO(2)$ is the rotational symmetry of $\mathbb{R}^2$ transverse to the defect.

The superconformal index of the infrared theory is defined in terms of the 3d $\mathcal{N}=2$ superconformal algebra. It is expanded by the fugacity $q$ related to the conformal dimension of the BPS operators. The fugacity $p$ of the UV index becomes the fugacity for $U(1)_f$ flavor symmetry. As the R-symmetry in the $\mathcal{N}=2$ superconformal group is the $U(1)_R$ abelian symmetry it can in general mix with other abelian flavor symmetries, in particular with $U(1)_f$ symmetry. In analogy with examples in higher codimension, 
it is likely that the correct R-symmetry of the low energy 3d/5d system may be determined by extremizing over the 3d flavor fugacities the $S^5$ partition function of the 5d CFT with the 3d defect wrapping an $S^3$ inside the $S^5$.

\section{Five- and three-dimensional systems from $(p,q)$ webs} \label{sec:webs}
The space of five-dimensional $\CN=1$ SCFTs is currently rather unexplored. 
A large class of examples can be built in string theory in terms of webs of fivebranes in IIB 
string theory, or equivalently toric Calabi-Yau singularities in M-theory. 

A $(p,q)$ web configuration can be labelled by a Newton polygon, a convex polygon 
with vertices located at integral points. To each edge $i$ of the polygon we can associate a vector of two integers, which we factorize uniquely as 
$n_i (p_i, q_i)$, with $p_i$ and $q_i$ relatively prime and $n_i>0$. This corresponds to a bundle of $n_i$ parallel semi-infinite 
$(p_i,q_i)$-fivebranes, with slope perpendicular to the edge. Polygons related by an $SL(2,\IZ)$ transformation on the plane 
(together with a IIB duality transformation) give equivalent systems. 

If all fivebranes converge to the origin, the string theory background is expected to give rise at low energy to some 5d SCFT
located at the origin, weakly coupled to six-dimensional $U(n_i)$ gauge theories located on the bundles of semi-infinite branes. 
Thus the 5d SCFT is expected to have an $\prod_i U(n_i)$ flavor symmetry, though a more careful inspection shows that 
some overall $U(1)$ factors act trivially. 

The 5d SCFTs have a Coulomb branches described as a generic resolution of the five-brane junction into 
a planar web with trivalent junctions. This is dual to a decomposition of the Newton polygon into triangles with vertices at integral points. 
Thus each integral point in the interior of the polygon is associated to a finite face of the web, whose size is a Coulomb branch modulus for the 
system. The transverse positions of the semi-infinite fivebranes are mass parameters for the $\prod_i U(n_i)$ flavor symmetries. 
Overall translations of the centre of mass of the webs give trivial $U(1)$ symmetries. Sometimes the transverse positions 
may be constrained geometrically, and the corresponding $U(1)$ symmetry is also lost. 

A particularly important class of five-brane web configurations corresponds to bundles of D5 (i.e. $(0,1)$ ) fivebranes suspended 
between NS5 branes (more precisely, $(1,q)$ fivebranes). Such a web gives a low-energy five-dimensional gauge theory description 
for the corresponding 5d SCFT, a linear quiver gauge theory with $SU(N_i)$ gauge groups, fundamental flavors at the end of the quiver only 
associated to semi-infinite D5 branes. These will be the main theories of interest for us. 

The five-dimensional SCFTs also have Higgs branches of vacua, or more precisely mixed Higgs-Coulomb branches. Higgs branch directions open up at
special loci on the Coulomb branch. There are at least two types of Higgs branch directions which are visible from the 
$(p,q)$ five brane construction: geometric transitions and seven-brane deformations. 

The first type of Higgs branch deformation simply corresponds to splitting the web into two parallel sub-webs,  
separated along one of the three directions perpendicular to the original web. Such a deformation is only available 
if a proper subset of the edges of the original Newton polygon adds up to zero. These Higgs branch deformations open up 
whenever the Coulomb branch parameters are tuned to reduce the original web to a superposition of the two sub-webs. 

These Higgs branches can be given a low-energy description in terms of extra light degrees of freedom appearing at the loci in the Coulomb branch where 
the web reduces to two intersecting sub webs. The extra degrees of freedom are associated to each intersection point between the two webs.
Elementary intersections between strands of type $(p,q)$ and $(p',q')$ with $p q' - q p' = \pm 1$ can be dualized to intersections of D5 and NS5 branes and thus 
give rise to standard hypermultiplets. We expect that general intersections between strands of type $(p,q)$ and $(p',q')$ with $p q' - q p' = \pm k$ give rise to $A_{k-1}$ singularities. 
This can be checked by comparing the Higgs branches for different phases of a 5d SCFT which admits a gauge theory description in some phase. 
We are assuming $(p,q)$ to be co-prime, and $(p',q')$ to be co-prime as well.  

Each intersection $a$ gives rise to some set of degrees of freedom equipped with a $U(1)$ flavor symmetry, whose three 
moment maps $\mu^I_a$ parameterize the separation of the two intersecting fivebranes. 
As the two sub-webs move rigidly in the transverse direction, all the moment maps must be equal to each other. 
This can be implemented by a hyper-K\"ahler quotient by the differences of $U(1)$ isometries associated to different intersections. 

These deformations will provide us with an easy way to implement 
Higgsing constructions: given a theory $T$ implemented by a fivebrane web, we can simply add to the web a new full 
fivebrane, i.e. two parallel edges $(p,q)$ and $(-p,-q)$, to get a larger theory $T'$ which a Higgs branch deformation flowing to $T$. 

The second type of Higgs branch deformation becomes visible if we decouple the six-dimensional gauge fields in the 
string theory description by having each of the semi-infinite fivebranes end on a seven brane of the same type, implementing Dirichlet 
boundary conditions. 

\begin{figure}[h]
    \centering
    \includegraphics[width=.99\textwidth]{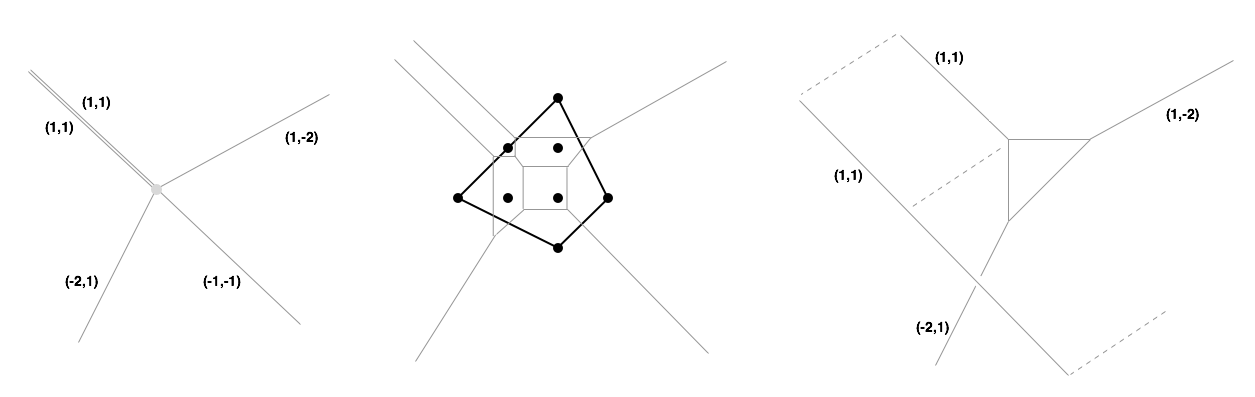} 
    \caption{Left: A generic fivebrane web, engineering an SCFT with an $SU(2) \times U(1)$ flavor symmetry. Center: the Newton polygon for the web and a generic Coulomb branch deformation. Right: A Higgs branch deformation 
    of the 5d SCFT. At low energy, we find a simpler SCFT, associated to the web with edges $(1,1)$, $(1,-2)$, $(-2,1)$. }
    \label{fig:two}
\end{figure}

As in lower dimensional examples, once the sevenbranes are added to the construction we can produce new 
theories by having several fivebranes of the same type end on the same sevenbrane. The resulting theories are labelled by an additional choice of 
$\mathfrak{su}(2)$ embeddings $\rho_i$ in $U(n_i)$, and can be found at the bottom of RG flows initiated from the Higgs branch deformations of the original class of theories. 
In a gauge theory context, these constructions give linear quiver gauge theories with fundamental flavors at generic nodes. 

\begin{figure}[h]
    \centering
    \includegraphics[width=.99\textwidth]{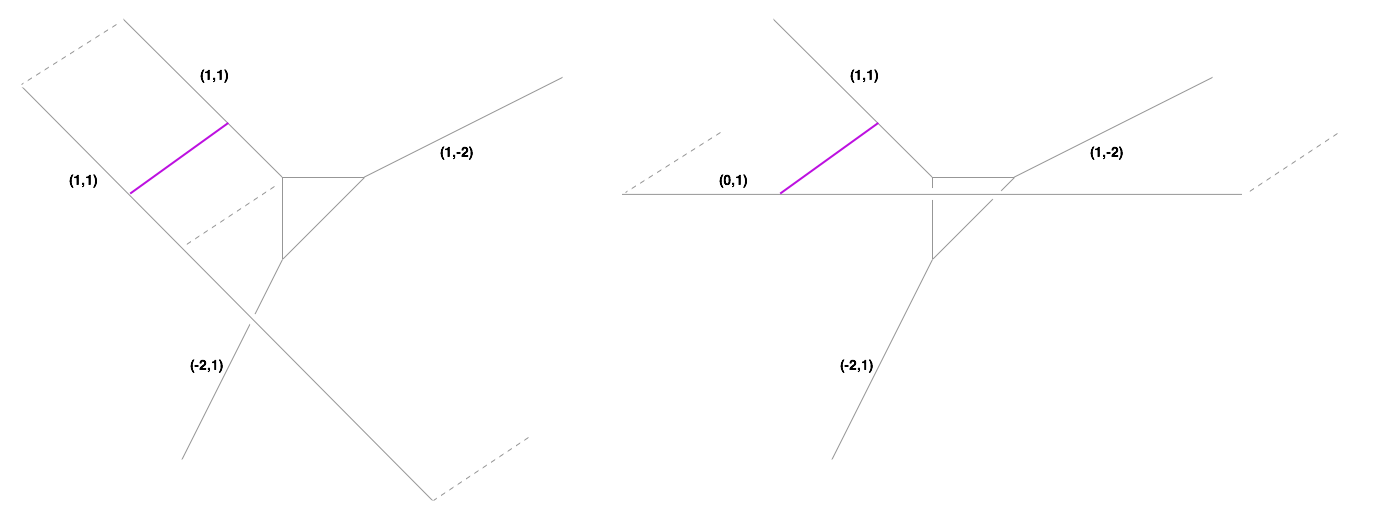} 
    \caption{A position dependent Higgs branch vev in the theory of Figure \ref{fig:two} produces a $D^{(1)}_{(1,1)}$ defect in the simpler SCFT. A position dependent Higgs branch vev in a different theory produces a $D^{(1)}_{(0,1)}$ defect in the simpler SCFT. The two defects should be related by a 3d $SL(2,\IZ)$ transformation. }
    \label{fig:three}
\end{figure}

Co-dimension two defects in these five-dimensional systems can be implemented in string theory by 
one or more D3 branes ending on the fivebrane web. More precisely, a D3 brane supports a free four-dimensional 
$U(1)$ $\CN=4$ gauge theory on a half space, coupled at a boundary to the three-dimensional degrees of freedom added to the five-dimensional SCFT. 
We can decouple the four-dimensional theory by having the D3 brane end on an extra $(p,q)$ fivebrane separate from the original system. 
This produces a family of co-dimension two defects $D^{(1)}_{(p,q)}$ in the five-dimensional SCFT, labelled by the pair $(p,q)$. 
The defects are equipped with a $U(1)$ flavor symmetry. Witten's $SL(2,\IZ)$ action~\cite{Witten:2003ya} on three-dimensional $\CN=2$ 
theories equipped with an $U(1)$ flavor symmetry acts on the $(p,q)$ labels in the obvious way. 

The construction of a co-dimension two defect from a D3 brane stretched between 
the fivebrane web and an extra $(p,q)$ fivebrane gives an immediate link to the Higgsing construction of
co-dimension two defects: the extra fivebrane converts the original theory $T$ into the extended
theory $T'$, far along the Higgs branch where it reduces back to $T$. The D3 brane represents a vortex 
configuration in that Higgs branch, which engineers the defect in field theory. 

On the other hand, if we have a system of fivebranes with a gauge theory interpretation, the D3 brane defect can be given one or more
GLSM descriptions: the 3d degrees of freedom arise from D3-D5 strings. This provides a bridge between 
constructions based on Higgsing and GSLM constructions. 

\subsection{Example: Free hypermultiplets}
The intersection of a single D5 brane and a single NS5 brane is known to support a 
single hypermultiplet. The intersection can be resolved in two ways to a web with either an $(1,1)$ or an $(1,-1)$ 
intermediate edges, the toric diagram of a conifold. These two webs correspond to the two triangulations of a $1 \times 1$ square. 
By $SL(2,\IZ)$ symmetry, a single hypermultiplet also arises at the intersection of any 
pair of fivebranes of type $(p,q)$ and $(p',q')$ with $p q' - q p' = \pm 1$. 

The resolution makes the hypermultiplet massive.
In M-theory, the massive hypermultiplet is visible as a M2 brane wrapping 
the $\IC P^1$ corresponding to the intermediate edges. In string theory, it is a $(1,1)$ or $(1,-1)$ string 
stretched between the intersection points. 

Thus the mass of the hypermultiplet is controlled by the relative displacement of the 
two semi-infinite D5 branes, which is the same as the relative displacement of the 
two semi-infinite NS5 branes. 
Conversely, a hypermultiplet vev corresponds to the separation of the D5 and NS5 in one of the three transverse directions.
The relative separation is actually controlled by the three moment maps for the hypermultiplet. 

The co-dimension two defects engineered by a semi-infinite D3 brane ending on the system can be described in a straightforward manner.
It is known that the D3-D5 strings give a pair of 3d chiral multiplets $q$, $\tilde q$, coupled to the bulk hypermultiplet by a cubic superpotential \cite{Hanany:1997vm}
\begin{equation}
W = \phi \tilde \phi M
\end{equation}
with $M$ being one of the two complex scalars inside the hypermultiplet, restricted to the defect. Which of the two scalars depends on the direction from which the $D3$ 
brane ends on the system. 

If the D3 ends on a second D5 brane separate from the system, this is all the 3d matter present at the defect. We can denote that defect as $D^{(1)}_{0,1}$. 
The relative position of the D5 branes in the plane of the web is the 3d real mass for the isometry which rotates the 3d chiral multiplets in opposite directions. 
A more general defect can be obtained by $SL(2, \IZ)$ transformations. For example, if the D3 brane ends on a second NS5 brane, 
we gauge the $U(1)$ flavor symmetry of $\phi$ and $\tilde \phi$ and obtain three-dimensional $\CN=2$ SQED with one flavor, and a superpotential coupling 
between $M$ and the meson operator $\phi \tilde \phi$. We can denote that defect as $D^{(1)}_{1,0}$.

Notice that the brane system which engineers the free hypermultiplet is actually invariant under an $SL(2,\IZ)$ S transformation, accompanied by a rotation by ninety degrees of the 
web. Thus the two co-dimension two defects $D^{(1)}_{0,1}$ and $D^{(1)}_{1,0}$ should be actually dual to each other. 
We can verify this by using the basic $\CN=2$ mirror symmetry. The $N_f=1$ SQED is equivalent to an XYZ model, i.e. three chirals $x$,$y$,$z$ with 
superpotential coupling $xyz$. The duality matches $z = \phi \tilde \phi$. 
Coupling that to the bulk hypermultiplet, we get a superpotential 
\begin{equation}
W = x y z + M z \ .
\end{equation}

As we saw before, a 3d chiral $z$ with linear coupling to the bulk hypermultiplet can be reabsorbed into the bulk hypermultiplet 
itself, up to a shift of fugacities. Thus we eliminate $z$ and re-write the superpotential as 
\begin{equation}
W = x y \tilde M \ ,
\end{equation}
where $M$ is the second complex scalar in the bulk hypermultiplet, restricted to the defect.
This is indeed the $D^{(1)}_{0,1}$ defect. 

\begin{figure}[h]
    \centering
    \includegraphics[width=.99\textwidth]{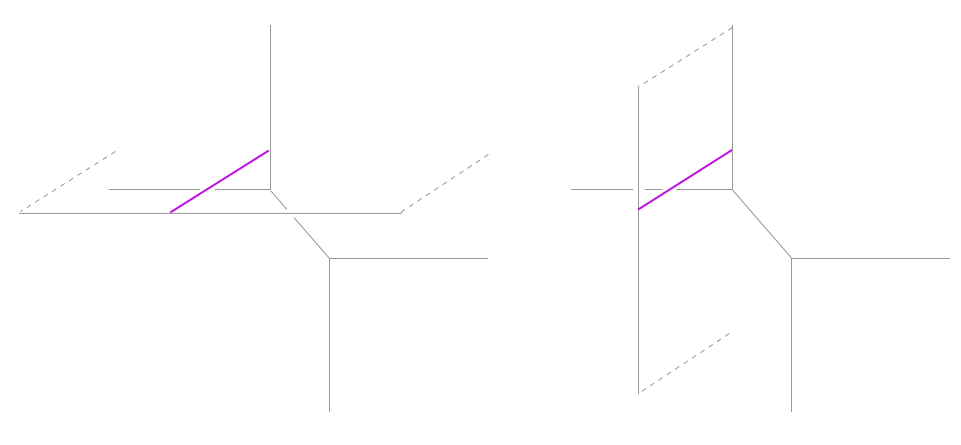} 
    \caption{The defects $D^{(1)}_{0,1}$ and $D^{(1)}_{1,0}$ in the hypermultiplet theory. They are related by a rotation of the plane by $\pi/2$, i.e. a 3d $S$ transformation}
    \label{fig:four}
\end{figure}

\subsection{Example: Pure $SU(2)$ gauge theory}
The simplest 5d SCFT which can be mass-deformed to a five-dimensional gauge theory 
arises at the intersection of an $(1,1)$ and an $(1,-1)$ fivebranes. By $SL(2,\IZ)$ symmetry, it also arises at the intersection of any 
pair of fivebranes of type $(p,q)$ and $(p',q')$ with $p q' - q p' = \pm 2$, such as a $(1,0)$ and a $(1,2)$ fivebrane. 
The theory is sometimes denoted as the $E_1$ theory. See \cite{Morrison:1996xf,Bergman:2013ala} for a discussion of several important features of the
SCFTs associated to the pure $SU(2)$ gauge theory. 

The intersection can be resolved into a square with sides made of D5 and NS5 brane segments, without 
moving the semi-infinite branes. Thus the 5d SCFT has a one (real) dimensional Coulomb branch, with the geometry of $\IR^+$. 
It also has a Higgs branch, corresponding to the transverse separation of the two five-branes. We will see momentarily that the Higgs branch 
has the geometry of an $A_1$ singularity. 

The theory has a mass deformation, corresponding to a relative displacement of two opposite semi-infinite fivebranes. 
A ``positive'' mass deformation $\ell$ gives us a system with two parallel D5 brane segments of finite length $\ell+\phi$ and arbitrary separation $\phi$
ending on $(1,k)$ branes, which support at low energy a pure $SU(2)$ gauge theory with gauge coupling $g_{YM}^{-2} = \ell$ and Coulomb branch parameter $\phi$. 
A negative mass deformation also gives a pure $SU(2)$ gauge theory with gauge coupling $g_{YM}^{-2} =- \ell$ and Coulomb branch parameter $\tilde \phi = \phi + \ell$, simply by acting with $S$ from $SL(2,\IZ)$. 

The $U(1)$ isometry associated to $\ell$, whose current is the instanton current of the 5d gauge theory, 
is known to be enhanced to $SU(2)$ in the 5d SCFT, and the symmetry $\ell \to - \ell$ is actually the Weyl symmetry.

For future reference, we may recall here the calculation of the index of this 5d SCFT. 
The index takes the following form:
\begin{equation}
I_{SU(2)}(\lambda;p,q) = \frac{I_{\mathrm{vec}}}{2} \oint \frac{da}{2 \pi i a} (a^\pm;p,q)_\infty (p q a^\pm;p,q)_\infty \left[\sum_{n=0}^\infty \lambda^n R_n(a;p,q)\right]\!\left[\sum_{n=0}^\infty \lambda^{-n} R_n(a^{-1};p,q)\right] \ .
\end{equation}
The sums in parentheses are the instanton partition function and its conjugate. They are defined as power series in the instanton counting parameter $\lambda$,
with coefficients which are rational functions $R_n(a;p,q)$ of $a$,$p$ and $q$. In order to compute the index, the series should be re-interpreted as series in the fugacities 
$p$ and $q$. The Weyl symmetry $\lambda \to \lambda^{-1}$ follows trivially from the expression of the index. A stricter test of symmetry enhancement 
is the presence of the moment map operators at order $p q$ in the index~\cite{Bashkirov:2012re}.

The instanton partition functions $Z_{\mathrm{inst}}^{SU(2)}$ in the formula above can be computed from Nekrasov's localization formulae for $U(2)$ (with Chern-Simons level $0$) or $Sp(1)$, 
or from the refined topological vertex.
It is interesting to observe that the refined topological vertex result is automatically symmetric under $\lambda \to \lambda^{-1}$ and $a \to \lambda a$, 
but only if considered as a series expansion at large $a$ and $\lambda$ of rational functions of $p$ and $q$, which is not equivalent to the 
series expansion relevant for index calculations.  

In anticipation of Higgsing calculations of defects $D^{(1)}_{p,q}$ corresponding to a single $D3$ brane stretched between the 
pure $SU(2)$ web and a $(p,q)$ fivebrane, 
we can describe the theories $T'$ which arise if we bring in contact the web for pure $SU(2)$ with the $(p,q)$ fivebrane. 

\begin{itemize}
\item For $D^{(1)}_{0,1}$, we obtain a web which is associated to an $SU(3)$ gauge theory with $N_f=2$. 
Separating the D5 brane from the web corresponds to a mesonic Higgs branch deformation. The $2 \times 2$ meson matrix  
$M = \tilde Q Q$ built from the fundamental $Q$ and anti-fundamental $\tilde Q$ chiral scalar fields in the flavor hypermultiplets 
satisfies the equation $M^2=0$, because of the F-term constraint $Q \tilde Q=0$. A nilpotent vev for $M$ Higgses $SU(3)$ back to $SU(2)$,
eating up all the hypers. 
\item For $D^{(1)}_{1,0}$, we obtain a web which is associated to a $SU(2) \times SU(2)$ quiver gauge theory, with no flavors besides the bi-fundamental 
hypermultiplet. Separating the NS5 brane corresponds to a baryonic Higgs branch deformation. The two bi-fundamental chiral fields 
$Q$ and $\tilde Q$ vev must satisfy F-term constraints $Q \tilde Q = \tilde Q Q =0$. The baryonic Higgs branch sets to zero, say, $\tilde Q$ 
and sets $Q$ to be an invertible $2 \times 2$ matrix, which we can take to be a multiple of the identity. This Higgses the product gauge group 
to the diagonal $SU(2)$, eating up all the hypers.  
\item For $D^{(1)}_{1,\pm 1}$, we obtain another web which is associated to a $SU(2) \times SU(2)$ quiver gauge theory, with no flavors besides the bi-fundamental 
hypermultiplet. Separating the  NS5 brane corresponds to the same baryonic Higgs branch as in the previous case. 
Thus we expect the $D^{(1)}_{1,\pm 1}$ defect to be essentially equivalent to the $D^{(1)}_{1,0}$ defect. This equivalence is not manifest 
in the brane system. This generalizes the 3d mirror symmetry relation between 3d $U(1)$ gauge theories coupled 
to a doublet of chiral fields and CS level $0$ or $\pm 1$, which can be obtained from a decoupling limit of the equivalence between defects.  
\end{itemize}

Before ending the discussion of the pure $SU(2)$ theory, it is useful to remember that there is a second $(p,q)$-fivebrane 
description of the same $E_1$ 5d SCFT, associated to semi-infinite five-branes with labels $(1,0)$, $(1,0)$, $(-1,2)$, $(-1,-2)$. 
The equivalence of the two descriptions is rather non-trivial, even if they both realize pure $SU(2)$ gauge theory at low energy. 
The parallel $(1,0)$ branes make the $SU(2)$ flavor symmetry of the 5d SCFT manifest: it is coupled to the six-dimensional $SU(2)$ 
gauge fields on the parallel fivebranes. The Higgs branch becomes manifest only if we end the semi-infinite $(1,0)$ branes on two 
$(1,0)$ sevenbranes, and it is described by the motion of the $(1,0)$ brane segment stretched between the sevenbranes. 
This type of brane configuration is known to give a Higgs branch with the geometry of an $A_1$ singularity. 

The instanton partition function can be computed from this second web realization in terms of a $U(2)$ instanton partition function with 5d CS term $2$,
as long as one removes a factor $(p q \lambda;p,q)_\infty$, which can be interpreted as being due to the presence of the two parallel NS5 branes in the second web realization. See~\cite{Bao:2013pwa,Hayashi:2013qwa} for more explanations of the decoupled factor in the low energy partition function.

Although the second web realizes the same 5d SCFT, if we add a D3 brane to the system 
we expect to get a different family of co-dimension two defects: after all, the webs control the geometry 
of the moduli space of vacua of the defects.

\begin{figure}[h]
    \centering
    \includegraphics[width=.99\textwidth]{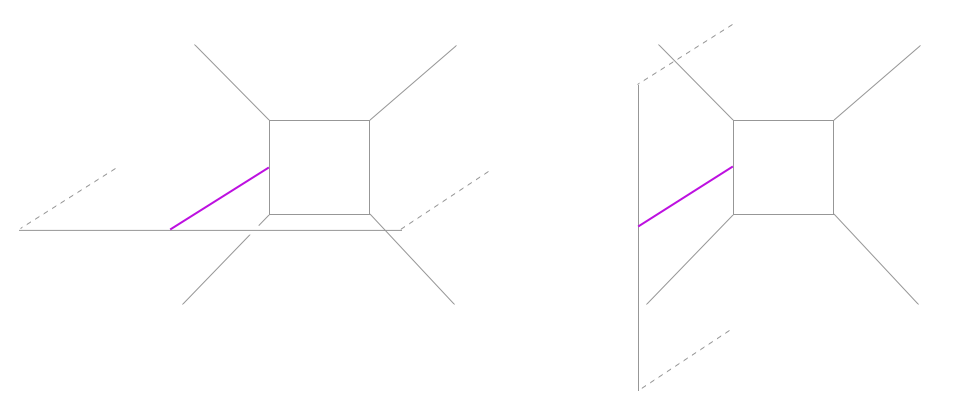} 
    \caption{The defects $D^{(1)}_{0,1}$ and $D^{(1)}_{1,0}$ in the $E_1$ (pure $SU(2)$) theory. They are related by a rotation of the plane by $\pi/2$, i.e. a 3d $S$ transformation}
    \label{fig:five}
\end{figure}

\subsection{Example: $SU(3)$ gauge theory with $N_f=2$}
We now engineer the co-dimension two defect of type $D_{0,1}^{(1)}$ from Higging of the UV theory $T'$. For this case, the theory $T'$ is the 5d SQCD with $N_c=3$ and $N_f=2$. The theory at the conformal fixed point is related to an intersection of three five-branes labelled by $(1,1),\, (1,-1),\, (0,1)$.  In the next section we shall explain the construction of the defect $D_{1,0}^{(1)}$ using a different UV theory.

The SQCD has a mesonic Higgs branch deformation realized by nontrivial vev of the meson matrix $M=\tilde{Q}Q$. The position dependent vev will lead to the pure $SU(2)$ gauge theory with a co-dimension two defect at the end of RG flow.
In the context of brane web diagram, the intersection of five-branes can be resolved to have two compact faces at the center, a square and a hexagon, corresponding to two dimensional Coulomb branch. The resolved web diagram has three finite D5 branes at the center realizing the $SU(3)$ gauge group and two semi-infinite D5 branes for the $N_f=2$ flavors. The Higgs branch opens up when the two semi-infinite D5 branes align with one of the finite $D5$ branes, which can form a single full $D5$ brane. The transverse separation of the full $D5$ brane is associated to the mesonic Higgs branch. The defect comes from a $D3$ brane connecting a NS5-brane to the transverse $D5$ brane.

The superconformal index of the 5d SCFT is
\begin{align}
&I^{3,2}_{\mathrm{SQCD}}(\mu_a,\lambda;p,q) \\
& = \frac{I_{\mathrm{vec}}^2}{3!} \oint \prod_{i=1}^3\frac{dz_i}{2\pi iz_i} \frac{ \prod_{i\neq j}^3 (z_i/z_j;p,q)_\infty (pqz_i/z_j;p,q)_\infty } { \prod_{i=1}^3 \prod_{a=1}^2 (\sqrt{pq}z_i/\mu_a;p,q)_\infty (\sqrt{pq}\mu_a/z_i;p,q)_\infty } \, \left|Z_{\mathrm{inst}}(z_i,\mu_a,-\lambda;p,q)\right|^2 \ .
\end{align}
The instanton partition function $Z_{\mathrm{inst}}$ can be computed from the Nekrasov's formulae for $U(3)$ (at CS-level 0) with 2 flavors. For later convenience we use the instanton fugacity $(-\lambda)$ instead of $\lambda$. The refined topological vertex on the relevant brane web diagram also computes the same partition function. 
The conjugation in $|\cdot|^2$ inverts all fugacities, including the instanton fugacity $\lambda$ as well as $p,q$.

The index has a simple pole at $\mu_1/\mu_2 p^{r+1}q^{s+1} = 1$ associated with the mesonic operator $M$ carrying angular momenta $r,s$.
If we take the residue at $\mu_1/\mu_2 p^{2}q = 1$, we get the index of the IR theory with a defect of type $D_{0,1}^{(1)}$. In brane description, it amounts to separating a full D5 brane away from the system and introducing a single D3 brane connecting the separated D5 brane and the rest of the system. 

The simple pole at $\mu_1/\mu_2 p^2q = 1$ arises when the two sets of poles at $(z_3 = \sqrt{pq}p\mu_1,z_3 = \sqrt{pq}^{-1}\mu_2)$ and $(z_3 = \sqrt{pq}\mu_1,z_3 = \sqrt{pq}^{-1}p^{-1}\mu_2)$ pinch the $z_3$ integral contour. Taking the residues at these poles we obtain the superconformal index of the IR theory with the $D_{0,1}^{(1)}$ defect:
\begin{align}
&I_{0,1}(\mu,\lambda;p,q) \equiv \lim_{\mu_1/\mu_2 p^2q \rightarrow 1} I_{\rm hyper}^{-1}(\sqrt{pq}\mu_1/\mu_2;p,q) \ I_{\mathrm{SQCD}}^{3,2}(\mu_a,\lambda;p,q)  \\
&= \frac{I_{\mathrm{vec}}}{2}  \oint \frac{da}{2\pi ia} (a^{\pm};p,q)_\infty(pqa^{\pm};p,q)_\infty I_{\mathrm{3dchiral}}(p^{-1/4}\sqrt{a}^\pm\mu) |Z^{0,1}_{\mathrm{inst}}(a,\mu,\lambda;p,q)|^2 + (\mu \rightarrow \mu^{-1}) \nonumber \,.
\end{align}
Here we have defined $a\equiv z_1/z_2, \mu \equiv (\mu_1\mu_2)^{3/4}$ and imposed the condition $z_1z_2z_3=1$. $Z^{0,1}_{\mathrm{inst}}(a,\mu,\lambda)$ is the instanton partition function with specialized fugacities at the pole $\mu_1/\mu_2p^2q=1$, which takes the form
\begin{align}
&Z^{0,1}_{\mathrm{inst}}(a,\mu,\lambda;p,q) \\
&= 1 + \lambda \left( \frac{\sqrt{p}q(1-p^{5/4}q/(\mu \sqrt{a}))}{(1-p)(1-q)(1-a)(1-p^{1/4}q/(\mu \sqrt{a}))(1-pq/a)} + (a\rightarrow a^{-1}) \right) + \mathcal{O}(\lambda^2)
\nonumber \ .
\end{align}
The first term in the superconformal index is the contribution from the first set of poles and the second term is from the second set. They can be interpreted as the contributions from the two fixed points of the vortex moduli space on the defect, i.e. the D3-brane ending on either of the NS5-branes. The factor $I_{\mathrm{3dchiral}}(p^{-1/4}\sqrt{a}^\pm\mu)$ is the contribution from the $SU(2)$ doublet of 3d chiral multiplets with the $U(1)$ flavor charge $+1$.



The index is expanded mainly by $q$ which is the fugacity for the conformal dimension of the superconformal algebra in the 5d/3d coupled system.
We note that the integrand in the index has some factors independent of the fugacity $q$ and thus one has to carefully evaluate the integral. We shall evaluate the integral by choosing a unit circle contour for $a$ with the assumption $q,p<\mu p^{-1/4}<1$. 
After evaluating the integral, we shall expand the final result by $q$ and $p$.
One can of course use a different assumption like $q,p<\mu^{-1} p^{-1/4}<1$, but the result turns out to be the same.
We obtain
\begin{equation}\label{eq-index-3dchiral-defect}
I_{0,1}(\mu,\lambda;p,q) = 1 - q +\sqrt{p}q(\lambda+1/\lambda)+(2p-q)q +\sqrt{p}q(p+q)(\lambda+1/\lambda) + \sqrt{p}q^2(\mu^2+1/\mu^2) +\cdots
\end{equation} We checked this result against the direct residue computation (after performing all the $SU(3)$ holonomy integrals first) and found agreement up to first few orders in $p,q$ expansion.

\subsection{Example: $SU(2)\times SU(2)$ gauge theory with a bi-fundamental}
The $D_{1,0}^{(1)}$ defect can be engineered by Higgsing an $SU(2)\times SU(2)$ gauge theory with a bi-fundamental hypermultiplet. The five-brane web construction for this gauge theory is related by S transformation to the five-brane web for the SQCD with $N_c=3$ and $N_f=2$ considered in the previous section. The S transformation corresponds to the S-duality in type IIB theory that rotates the plane of the $(p,q)$ web by $\pi/2$ which exchanges D5 and NS5 branes.
The theory admits a baryonic Higgs branch parametrized by the vev of the chiral operator formed by bi-fundamental scalar field $Q$. It takes the form $\epsilon^{ab}\epsilon_{\dot{a}\dot{b}}Q_a^{\dot{a}}Q_b^{\dot{b}}$ where $a,\dot{a}$ are $SU(2)$ doublet indices.

From the brane web perspective, the Higgs branch corresponds to removing a full NS5 brane from the five-brane web along the transverse direction. We can insert a D3-brane connecting the transverse NS5 brane and either of D5 branes in the middle of the ($p,q$) web. This D3 brane is the brane realization of the $D_{1,0}^{(1)}$ defect. Clearly, this system is related by S transformation to the 5d/3d coupled system with the $D_{0,1}^{(1)}$ defect. The 3d theory on the defect has a UV description as a $U(1)$ gauge theory with two chiral fields of flavor charge $+1$ which transform as a doublet under the bulk $SU(2)$ gauge symmetry. The theory has two massive vacua and they are mapped to the D3 brane configuration ending on either D5 brane.

The superconformal index of the 5d $SU(2)\times SU(2)$ quiver theory is given by
\begin{align}
&I_{2\times2}(\mu,\lambda_1,\lambda_2;p,q) \\
&= \frac{I_{\mathrm{vec}}^2}{4}\oint \frac{dadb}{(2\pi i)^2ab} \frac{(a^\pm;p,q)_\infty(pqa^\pm;p,q)_\infty(b^\pm;p,q)_\infty(pqb^\pm;p,q)_\infty}{(\sqrt{pq}\sqrt{a}^\pm \sqrt{b}^\pm \mu^\pm;p,q)_\infty} \big|Z_{\mathrm{inst}}^{2\times2}(a,b,\mu,\lambda_1,\lambda_2;p,q)\big|^2 \nonumber \ .
\end{align}
where $\mu$ is the fugacity for the $SU(2)$ flavor symmetry of the bi-fundamental matter.
The instanton partition function $Z_{\mathrm{inst}}^{2\times2}$ is expanded by instanton fugacities $\lambda_1$ and $\lambda_2$. We can compute the instanton partition function using the Nekrasov's localization formulae for the $U(2)\times U(2)$ gauge theory at CS-levels $(0,0)$ or the topological vertex on the five-brane web. 

The index has the baryonic branch poles at $\mu^{\pm2} p^{r+1}q^{s+1} = 1$. For the $D_{1,0}^{(1)}$ defect, we take the residue at the pole $\mu^2 = p^2q$ that can arise from two sets of poles pinching the $b$ contour: $(\sqrt{b}=\sqrt{a}/\mu \sqrt{p^3q},\sqrt{b}=\sqrt{a}\mu /\sqrt{pq})$ and $(\sqrt{b}=\sqrt{a}/\mu \sqrt{pq},\sqrt{b}=\sqrt{a}\mu /\sqrt{p^3q})$. Summing over the residues we obtain the superconformal index with the defect $D_{1,0}^{(1)}$:
\begin{align}
&I_{1,0}(\lambda_1,\lambda_2;p,q) =  \lim_{\mu^2 \rightarrow p^2q}
I_{\rm hyper}^{-1}(\sqrt{pq}\mu^{-2};p,q)\ I_{2\times2}(\lambda_1,\lambda_2;p,q) \\
&= I_{\mathrm{vec}}\oint \frac{da}{2\pi ia} \left[(a^\pm;p,q)_\infty(pqa^\pm;p,q)_\infty I_{\mathrm{3dchiral}}(a;q) \left|Z^{1,0}_{\mathrm{inst}}(a,\lambda_1,\lambda_2;p,q)\right|^2 + (a\rightarrow a^{-1})\right] \nonumber \ .
\end{align}
Here $Z^{1,0}_{\mathrm{inst}}(a,\lambda_1,\lambda_2)$ is the instanton partition function with specialized fugacities at each pole.
The first few terms in the instanton partition function are
\begin{equation}
Z^{1,0}_{\mathrm{inst}} = 1+\lambda_1\frac{\sqrt{p}q\sqrt{a}}{(1-q)(1-pqa)} + \lambda_2\frac{q/\sqrt{a}}{(1-q)(1-q/a)} + \lambda_1\lambda_2 \frac{\sqrt{p}q(1-pq^2)}{(1-p)(1-q)^2(1-q/a)(1-pqa)} + \cdots
\end{equation}
Two contributions in the integrand related by $a\leftrightarrow a^{-1}$ can be understood as the contributions from the two vacua of the 3d theory on the defect.

We take the unit circle contour for $a$ and evaluate the integral. The index is then expressed as a series expansion in terms of the parameters $p$ and $q$ as follows:
\begin{align}
\hspace{-0.5cm} I_{1,0}(\lambda_1,\lambda_2;p,q) &= 1- q +\sqrt{p}q(\lambda_1\lambda_2+(\lambda_1\lambda_2)^{-1}) + (2p-q)q  \nonumber \\ & \quad +\sqrt{p}q(p+q)(\lambda_1\lambda_2+(\lambda_1\lambda_2)^{-1}) +\sqrt{p}q^2(\lambda_1/\lambda_2+\lambda_2/\lambda_1) + \cdots
\end{align}
One may notice that this index is the same index as that of the $D_{0,1}^{(1)}$ defect case given in (\ref{eq-index-3dchiral-defect}) upon the following identification
\begin{equation}
\lambda = \lambda_1\lambda_2 \,, \quad \mu^2 = \lambda_1/\lambda_2 \ .
\end{equation}
This is already expected. The $D_{0,1}^{(1)}$ and the $D_{1,0}^{(1)}$ defects are related by a reflection of the brane system along the diagonal, 
i.e. the Weyl symmetry of the enhanced $SU(2)$ flavor symmetry of the pure $SU(2)$ theory, which is broken to $U(1)$ by the defect. 
The same reflection also relates, of course, the fivebrane diagrams for the $SU(3)$ with $N_f=2$ flavors and the $SU(2) \times SU(2)$ quiver, which are two phases of the same 5d SCFT
and thus have the same index \cite{Bergman:2013aca}. 

\subsection{The S transformation}
As discussed earlier, $SL(2,\IZ)$ transformations of the brane system should act on $D_{p,q}^{(1)}$ defects as Witten's $SL(2,\IZ)$ action 
on 3d CFTs with an $U(1)$ global symmetry. In particular, the $S$ transformation should interchanges two different types of defect $D_{1,0}^{(1)}$ and $D_{0,1}^{(1)}$. 
The S transformation effectively gauges the $U(1)$ flavor symmetry and add a mixed CS term of the form $A_{\rm new} dA_{\rm old}$, or a FI term for the new gauge group.
We will look at how the S transformation is realized at the level of hemisphere partition functions and then superconformal indices in turn.

Let us first consider the S transformation on the 3d hemisphere partition functions or 3d holomorphic blocks \cite{Beem:2012mb}.
The 3d hemisphere partition function with various boundary conditions was recently computed in \cite{Yoshida:2014ssa}. 
The S transformation acts as
\begin{equation}
Z^{\mathrm{3d}}_{S^1\times \mathbb{R}^2_q}(\mu) \ \stackrel{S}{\longrightarrow} \ \int\frac{d\mu}{\mu} \frac{\theta(x;q)\theta(\mu;q)}{\theta(x\mu;q)} Z^{\mathrm{3d}}_{S^1\times \mathbb{R}^2_q}(\mu) \ .
\end{equation}
As the $U(1)$ flavor symmetry is gauged, the fugacity $\mu$ becomes the dynamical gauge fugacity and thus we have the integral over $\mu$. The theta functions can be interpreted as the elliptic genus of the boundary 2d theory which consists of two fermi and one chiral multiplets coupled to the new gauge symmetry. This boundary theory is introduced to cancel the gauge anomaly due to the background mixed CS term in the presence of the boundary.

In the decoupling limit $\lambda\rightarrow0$, the theory on the $D_{0,1}^{(1)}$ defect reduces to two free chirals and its partition function takes the form
\begin{equation}
Z^{\mathrm{3d}}_{D_{0,1}^{(1)}} = (p^{-1/4}\sqrt{a}^\pm\mu;q)_\infty^{-1}
\end{equation}
or, by multiplying $\theta(p^{-1/4}\sqrt{a}^\pm\mu;q)$,
\begin{equation}
\tilde{Z}^{\mathrm{3d}}_{D_{0,1}^{(1)}} = (qp^{1/4}\sqrt{a}^{\pm 1}\mu^{-1};q)_\infty \ .
\end{equation}
The former expression is with Neumann boundary condition while the latter is with Dirichlet boundary condition~\cite{Yoshida:2014ssa}.

We take the former expression and perform the S transformation with respect to $\mu$.
The integral over $\mu$ can be evaluated by taking residues at the poles arising from the chiral doublets, i.e. poles at $\mu = \sqrt{a}p^{1/4}q^{-n}$ with $n \ge 0$.
The contour integral yields
\begin{equation}
-\frac{\theta(x;q)\theta(\sqrt{p/a};q)}{\theta(x\sqrt{p/a};q)}\frac{1}{(q;q)_\infty(a;q)_\infty} \sum_{k=0}^\infty (-x)^{-k}\frac{q^{\frac{k(k+1)}{2}}}{(q/a;q)_k(q,q)_k} \ .
\end{equation}
This is the vortex partition function for the 3d theory in the decoupling limit of the $D_{1,0}^{(1)}$ defect. The parameter $-x^{-1}$ becomes the FI parameter of the $U(1)$ gauge group and $k$ becomes the vorticity of the 3d theory.

We can promote the S transformation between 3d theories to the S transformation in 3d/5d coupled systems. As for the analogue of three dimensions, the S transformation is implemented by gauging the 3d $U(1)$ flavor symmetry on the defect. 

In analogy with the 3d case, we can first work with an ``hemisphere index'' $II^{\mathrm{Dir}}(a,\lambda;p,q)$, arising from pure $SU(2)$ gauge theory with Dirichlet b.c.
The hemisphere index simply consists of a holomorphic half of the full index integrand, with no integral done on the gauge fugacity $a$, 
which plays the role of the boundary global symmetry associated to Dirichlet boundary conditions. We do not know if such a boundary condition actually makes sense 
in the full 5d UV SCFT. The hemisphere index in the absence of defects can be written as 
\begin{equation}
II(a,\lambda;p,q) = (pq;p,q)_\infty(pqa^\pm;p,q)_\infty Z_{\mathrm{inst}}(a,\lambda;p,q)
\end{equation}
The expression for the hemisphere index is very simply understood: the full sphere index can be recovered from two hemispheres by gauging the diagonal 
boundary flavor symmetry of the Dirichlet boundary conditions by an $N=1$ vector multiplet

For the $D_{0,1}^{(1)}$ defect, we can define the hemisphere index as \begin{equation}
II_{0,1}(a,\mu,\lambda;p,q) = (pq;p,q)_\infty(pqa^\pm;p,q)_\infty (p^{-1/4}\sqrt{a}^{\pm}\mu;q)_\infty^{-1}Z_{\mathrm{inst}}^{0,1}(a,\mu,-\lambda;p,q)
\end{equation}
for the first fixed point/holomorphic block. The hemisphere index for the second fixed point/holomorphic block can be similarly defined by the replacement $\mu\rightarrow \mu^{-1}$. We can also define the hemisphere index for the $D_{1,0}^{(1)}$ defect such as
\begin{equation}
II_{1,0}(a,\lambda_1,\lambda_2;p,q) = (pq,p,q)_\infty(pqa^\pm;p,q)_\infty (qa^{-1};q)_\infty \, Z_{\mathrm{inst}}^{1,0}(a,\lambda_1,\lambda_2;p,q)
\end{equation}
for the first fixed point and similarly for the second fixed point by replacing $a\rightarrow a^{-1}$.
We have omitted the prefactors coming from regularizing the infinite products. We will ignore these prefactors in what follows as we are mainly interested in the superconformal index for which the prefactors are cancelled.

The $S$ transformation acts on the hemisphere partition function as
\begin{equation}
II_{0,1}(a,\mu,\lambda;p,q)	\ \stackrel{S}{\longrightarrow} \ \oint \frac{d\mu}{\mu}\frac{\theta(x;q)\theta(\mu;q)}{\theta(x\mu;q)} II_{0,1}(a,\mu,\lambda;p,q) \ .
\end{equation}
The contour integral is rather complicated than the 3d cases because not only the perturbative contribution but also the instanton series develops nontrivial poles. We suggest the contour prescription such that the contour encloses the poles at $\mu = \sqrt{a}p^{1/4}q^{-n}$ with any integer $n$. Note that the perturbative contribution has poles for $n\ge 0$ whereas $k$ instanton contribution has poles at $0< n \le k$. 

We compute the contour integral order by order in $\lambda$ and $x$ expansion and find that
\begin{align}
&\oint \frac{d\mu}{\mu}\frac{\theta(x;q)\theta(\mu;q)}{\theta(x\mu;q)} II_{0,1}(a,\mu,\lambda;p,q) \cr
&= - \frac{\theta(x;q)\theta(\sqrt{p/a};q)}{\theta(x\sqrt{p/a};q)(q;q)_\infty}\frac{(pq,p,q)_\infty(pqa^\pm;p,q)_\infty}{(a;q)_\infty} \, Z^{1,0}_{\mathrm{inst}} (a,\lambda_1 = \sqrt{a}^{-1}x\lambda,\lambda_2=\sqrt{a}x^{-1};p,q) \cr
&= - \frac{\theta(x;q)\theta(\sqrt{p/a};q)}{\theta(x\sqrt{p/a};q)(q;q)_\infty} II_{1,0}(a,\lambda_1 = \sqrt{a}^{-1}x\lambda,\lambda_2=\sqrt{a}x^{-1};p,q) \ .
\end{align}
This exhibits explicitly that the $S$ transformation maps the hemisphere partition function of the $D_{0,1}^{(1)}$ defect to the partition function of the $D_{1,0}^{(1)}$ defect,
up to multiplicative theta functions which are common in the $SL(2,\IZ)$ transformations of holomorphic blocks.
One can also take another contour enclosing the poles at $\mu=\sqrt{a}^{-1}p^{1/4}q^{-n}$ instead and get the result for the second fixed point of the $D_{1,0}^{(1)}$ defect.

We now turn to the S transformation at the level of superconformal index. In three dimensions, $SL(2,\IZ)$ transformations become transparent in the charge basis of the index, where the electric and magnetic charges for the flavor symmetry are fixed \cite{Dimofte:2011py}. The S transformation simply exchanges the electric and magnetic charges.
Likewise, we find that the S transformation in five dimensions also exchanges the electric and magnetic charges of the flavor symmetry in the superconformal index.

To see this, we need to figure out how to introduce the background magnetic flux for the $U(1)$ flavor symmetry and go to the charge basis. 
We describe the details of our proposal in the next section, and only report the result here. 
The index of the $D_{0,1}^{(1)}$ defect in the charge basis can be written as
\begin{align}
&I_{0,1}(e,m,\lambda) \\
&=\oint \frac{d\mu}{2\pi i\mu} \mu^{-e}  \Bigg[ \ \frac{1}{2}  \oint \frac{da}{2\pi ia} (a^\pm;p,q)_\infty(pqa^\pm;p,q)_\infty  \ \times \cr
&\quad   (p^{-1/4}\mu/\sqrt{q\lambda})^m\frac{(q^{\frac{2-m}{2}}p^{1/4}\sqrt{a}^\pm/\mu;q)_\infty}{(q^{-\frac{m}{2}}p^{-1/4}\sqrt{a}^\pm\mu;q)_\infty} \left|Z^{0,1}_{\mathrm{inst}}(a,\mu q^{\frac{m}{2}},\lambda;p,q)\right|^2 + (\mu,m) \rightarrow (1/\mu,-m) \Bigg] \nonumber \ .
\end{align}
The factor $(p^{-1/4}\mu/\sqrt{q})^m$ can be understood as a zero point shift of the charges in the background magnetic flux $m$. 
The $\mu$ integral is over a unit circle and it would be performed after the gauge holonomy integral so that it projects onto the states of the $U(1)$ flavor charge $e$. The contour for the gauge holonomy $a$ is taken to be a unit circle and the integral is performed with the assumption $q < \mu p^{1/4}< 1$.

Similarly, the charge basis index of the $D_{1,0}^{(1)}$ defect can be written as
\begin{equation}
I_{1,0}(e,m,\lambda) = \oint \frac{d\mu}{2\pi i \mu} \mu^{-e} \ I_{1,0} ( \lambda_1=\lambda \mu ,\lambda_2 = \lambda/\mu, m) \ .
\end{equation}

We find that two superconformal indices are related by the S transformation.
\begin{equation}
I_{1,0}(e,m,\lambda) = I_{0,1}(-m,e,\lambda) \ .
\end{equation}
Furthermore, it turns out that these indices are self-dual, namely
\begin{equation}
I_{0,1}(e,m,\lambda)=I_{0,1}(m,e,\lambda) \,, \quad I_{1,0}(e,m,\lambda)=I_{1,0}(m,e,\lambda) \ .
\end{equation}
We have checked these relations for $m,e=0,\pm1,\cdots,\pm4$ up to order $\mathcal{O}(p^{n_1}q^{n_2})$ with $n_1+n_2 < 6$. 

\begin{figure}[t]
\begin{center}
\includegraphics[width=100mm]{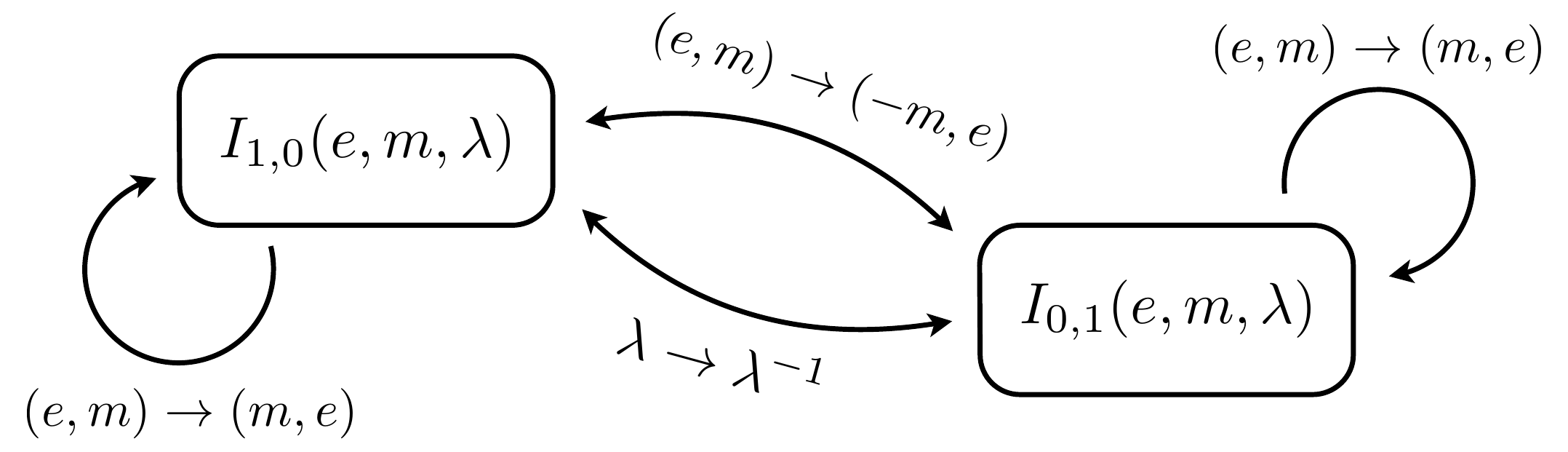}
\end{center}
\vspace{-20pt}
\caption{Dualities in superconformal indices  }
\label{fig:linear}
\end{figure}

The symmetries, depicted in Figure \ref{fig:linear}, reflect the expected geometric symmetries of the brane system, depicted in Figure \ref{fig:five}. 

\section{Difference equations and Wilson loops} \label{sec:diff}
\subsection{Wilson loops}
Another interesting objects we can add to the 3d/5d system are BPS Wilson loops. In general, one can add Wilson loops at the North and South poles of the sphere, wrapping the time circle. Their localized contributions to the superconformal index can be obtained from the Wilson loop partition function on $S^1\times \mathbb{C}^2$.
At first we consider for simplicity the hemisphere index with Wilson loops.
 The BPS Wilson loops are placed at the origin of $\mathbb{C}^2$ and wrap the time circle. When the Wilson loop is in the fundamental representation of the gauge group, we can interpret it as a heavy external fundamental string ending on the dynamical D5 branes in the brane picture. 

The hemisphere index with Wilson loops can be computed using the localization. However, the computation on the instanton background is rather nontrivial. Though the ADHM construction of the instanton moduli space is not affected in the presence of Wilson loops, but the Wilson loops add an additional vector bundles on the instanton moduli space. We should take into account this effect.

We need to know the equivariant Chern character for the additional vector bundle to compute the Wilson loop contribution.
For the fundamental Wilson loop, the vector bundle added is called universal bundle $\mathcal{E}$. The relevant equivariant Chern character at $k$ instantons is~\cite{Losev:2003py,Shadchin:2004yx}
\begin{equation}
Ch_{\mathcal{E}}(z_i,\rho_I;p,q) = \sum_{i=1}^N z^i -(1-p)(1-q)\sqrt{pq}^{-1} \sum_{I=1}^k \rho_I \ .
\end{equation}
Plugging this into the equivariant localization, we compute the hemisphere index with the Wilson loop in the fundamental representation
\begin{equation}
	\langle W_{\mathrm{fund}} \rangle = \frac{\sum_k \lambda^k \oint \prod_I d\rho_I \ Ch_{\mathcal{E}}(z_i,\rho_I;p,q) \ Z_k(z_i,\rho_I;p,q)}{\sum_k \lambda^k \oint \prod_I d\rho_I \ Z_k(z_i,\rho_I ;p,q)} \ ,
\end{equation}
which is normalized by the bare hemisphere index.

We are interested in the Wilson loop partition function in the presence of the co-dimension two defects. In order to compute this we start from the Wilson loops in the UV theory and apply the standard Higgsing procedure, which leads to the Wilson loops in the infrared theory with defects.
The partition function in the IR theory can be obtained from the UV partition function by taking residues at the poles corresponding to the defect. However, the normalized Wilson loop partition function of the UV theory does not have the pole as they are already cancelled through the normalization. Therefore we simply specialize the fugacities to the values at the poles for the defect.

For example, after Higgsing, we find that the fundamental Wilson loop partition function of the UV SQCD with $N_c=3$ and $N_f=2$ is related to that of the IR theory with the $D_{0,1}^{(1)}$ defect such as
\begin{equation}
	\langle W_{\mathrm{fund}} \rangle_{\mathrm{SQCD}}^{3,2} \stackrel{Higgsing}{\rightarrow} \sqrt{z_3}^{-1} \langle W_{\mathrm{fund}} \rangle_{D_{0,1}^{(1)}} + z_3 \ ,
\end{equation}
where $z_3$ takes the pole value for the defect. If one Higgs the UV theory by setting $\mu_1/\mu_2p^2q=1$ and $z_3=\sqrt{p}\mu$, one can obtain the Wilson loop partition function of the IR theory at one of two fixed points. The result is expanded as 
\begin{equation}
\langle W_{\mathrm{fund}} \rangle_{D_{0,1}^{(1)}}=\sqrt{a}+\sqrt{a}^{-1}-\lambda \left( \frac{\sqrt{p}q/\sqrt{a}(1-p^{3/2}q/(\sqrt{a}\mu))}{(1-a)(1-pq/a)(1-\sqrt{p}q/(\sqrt{a}\mu))} +(a\rightarrow a^{-1}) \right) + \cdots
\end{equation}

For the $D_{1,0}^{(1)}$ defect, the UV theory can have a fundamental Wilson loop in either gauge groups $SU(2)\times SU(2)$. We shall add a Wilson loop for the first gauge group whose Coulomb branch parameter is not tuned in the Higgsing procedure. After Higgsing, the Wilson loop in the UV theory reduces to the Wilson loop of the theory with the $D_{1,0}^{(1)}$ defect
\begin{equation}
	\langle W_{\mathrm{fund}} \rangle_{2\times2} \stackrel{Higgsing}{\rightarrow} \langle W_{\mathrm{fund}} \rangle_{D_{1,0}^{(1)}} \ .
\end{equation}
For example one can compute the contribution of one of the two fixed points to the Wilson loop partition function by setting $\mu^2=p^2q$ and $b=ap$ in the UV index. The result is expanded as
\begin{equation}
\langle W_{\mathrm{fund}} \rangle_{D_{1,0}^{(1)}}= \sqrt{a}+\sqrt{a}^{-1} - \lambda_1\frac{\sqrt{p}qa(1-p)}{(1-pqa)} - \lambda_1\lambda_2 \frac{\sqrt{p}q\sqrt{a}^{-1}(1+a)}{(1-q/a)(1-pqa)} + \cdots
\end{equation}

We can also consider the superconformal index with Wilson loops inserted at the North or South pole. The above hemisphere index with Wilson loops can be used here.
Let us denote by $W^{\rm hemi}$ the hemisphere index with Wilson loops.
The insertion of the fundamental Wilson loop, at the North pole for example, in the superconformal index amounts to inserting $W^{\rm hemi}$ into the gauge holonomy integral:
\begin{equation}
\langle W \rangle^{\mathrm {SCI}} = \frac{\oint \big[ \cdots W^{\rm hemi} \big]}{ \oint \big[ \cdots \big] } \ ,
\end{equation}
where $\cdots$ stands for the measure of the superconformal index without Wilson loop.
The superconformal index with a Wilson loop at the south pole then can be obtained by inserting into the integrand the complex conjugate of $W^{\rm hemi}$.

\subsection{Difference equation}
The partition functions of three-dimensional SUSY field theories on a hemisphere $S^1 \times \mathbb{C}_{q}$, where $q$ stands for the $\epsilon$ deformation, is known to be solutions to certain difference equations~\cite{Beem:2012mb}. The difference equations are interpreted as Ward-Takahashi identities for line operators located at the center of $\mathbb{C}_q$. In the classical limit $q\rightarrow 1$, it becomes a algebraic curve describing moduli space of the supersymmetric parameters of the 3d theory. This curve is a complex Lagrangian sub-manifold parametrized by fugacities $x$ for flavor symmetries and their momentum conjugates $p_x$. The deformation by $q$ amounts to the quantization of this algebraic curve and promotes the coordinates to the non-commuting operators obeying the relation $p_xx = qxp_x$.

We expect a similar correspondence to hold for co-dimension two defects in five-dimensional theories. In four dimensions, the Seiberg-Witten curve for a gauge theory can be interpreted as describing the moduli space of vacua of a co-dimension two defect, 
say a D2 defect for a gauge theory with a brane engineering construction~\cite{Alday:2009fs,Gaiotto:2013sma}. The Seiberg-Witten curve is promoted to a BPZ-like differential 
equation satisfied by the instanton partition function or the $S^4_b$ partition function. The differential equation does not always take a closed form: it encodes a chiral ring relation between twisted chiral operators on the defect and bulk Coulomb branch operators, both inserted in the 
instanton partition function. Sometimes these operator insertions can be traded for derivatives with respect to the couplings. Sometimes they cannot. 
The existence of BPZ-like differential equations is closely related to certain $qq$-character relations in the instanton partition function \cite{N2014}.

In a 3d/5d system we expect a similar story, upon compactification on a circle. The Seiberg-Witten curve for the 
compactified five-dimensional gauge theory corresponds to the moduli space of vacua of a co-dimension two defect, 
say $D^{(1)}_{(p,q)}$ for a theory with a brane web construction. Thus we expect the 5d instanton partition function or index to satisfy 
some type of difference equation, which encodes a 3d version of the twisted chiral ring relations. The difference operators 
in the 3d masses or FI parameters can be interpreted as the effect of inserting simple 1d line defects localized on the 3d defect. 
We do not expect, though, to be able to trade the insertion of bulk line defects for difference operators acting on the bulk data. 
Thus the difference equations will relate instanton partition functions or indices with or without Wilson loop insertions. 

 We shall check that the hemisphere index with co-dimension two defects satisfies a difference equation. The first example is the partition function with a $D_{0,1}^{(1)}$ defect. 
We find that the hemisphere index $II_{0,1}$ is annihilated by the following difference equation
\begin{equation}
\sqrt{a}p_\mu +\sqrt{p/a}\lambda p_\mu^{-1} + (p^{-1/4}\mu + p^{1/4}\mu^{-1}) =  W_{\mathrm{fund}} \ ,
\end{equation}
with the conjugate momentum $p_\mu$ of the $U(1)$ flavor fugacity $\mu$, where $W_{\mathrm{fund}}$ represents the insertion of the Wilson loop in the fundamental of the bulk gauge group. In the decoupling limit when $\lambda\rightarrow0$, the equation reduces to the known difference equation for the 3d theory with two chiral multiplets:
\begin{equation}
\sqrt{a}p_\mu  + (p^{-1/4}\mu + p^{1/4}\mu^{-1}) = \sqrt{a}+ \sqrt{a}^{-1} \ .
\end{equation}
It is instructive to redefine the partition function as
\begin{equation}
\mathcal{II}_{0,1}^{\mathrm{Dir}} = \frac{\theta(\mu;q)}{\theta(\mu /\sqrt{a};q)}II^{\mathrm{Dir}}_{0,1}
\end{equation}
and rewrite the corresponding difference equation as
\begin{equation}
p_{\tilde\mu} + \tilde\lambda p_{\tilde\mu}^{-1} + \tilde\mu + \tilde{\mu}^{-1} =  W_{\mathrm{fund}} \ ,
\end{equation}
with new parameters $\tilde\mu \equiv p^{-1/4}\mu$ and $\tilde\lambda \equiv \sqrt{p}\lambda$. In the Nekrasov-Shatashvili limit, when $p\rightarrow 1$, this equation becomes the quantum hamiltonian of 2-body closed Toda system. The partition function with the defect becomes the eigenfunction of the hamiltonian and the Wilson loop expectation value becomes the eigenvalue.

The second example is the hemisphere partition function with a $D_{1,0}^{(1)}$ defect.
We find that the hemisphere index $II_{1,0}$  obeys the following difference equation:
\begin{equation}
\sqrt{a}p_\tau + \sqrt{a}^{-1}p_\tau^{-1} - (\sqrt{p} \tau + \lambda\tau^{-1}) =  W_{\mathrm{fund}} \ ,
\end{equation}
where $\tau \equiv \lambda_1 ,\, \lambda \equiv \lambda_1\lambda_2$ and $p_\tau$ is the conjugate momentum of $\tau$. One may notice that this difference equation resembles the difference equation for the $D_{0,1}^{(1)}$ defect once we exchange the position and momentum variables, $p_\mu \leftrightarrow \tau$ and $p_\tau \leftrightarrow \mu$. This becomes clear if we redefine the partition function as
\begin{equation}
\mathcal{II}^{\mathrm{Dir}}_{1,0} = \frac{\theta(\tau;q)}{\theta(\tau \sqrt{a};q)}II^{\mathrm{Dir}}_{1,0}
\end{equation}
and the fugacities as $\tilde\tau \equiv -\sqrt{p}\tau$ and $\tilde\lambda \equiv \lambda\sqrt{p}$. Then it satisfies the following difference equation:
\begin{equation}
p_{\tilde\tau} + p_{\tilde\tau}^{-1}+\tilde\tau + \tilde\lambda\tilde\tau^{-1} = W_{\mathrm{fund}} \ ,
\end{equation}
Note that, after exchanging the position and momentum variables, this becomes the same difference equation as that of the defects $D_{1,0}^{(1)}$.
Of course, two defects and two difference equations are related by the S transformation discussed in the previous section. 

In the Nekrasov-Shatashvili limit, the difference equation becomes the hamiltonian of the integrable system which is bi-spectral dual of the closed Toda system and the partition function again becomes the eigenfunction of the hamiltonian.

The difference equations can also be understood in the context of the superconformal index. For doing so we need to introduce the background magnetic monopoles flux for the $U(1)$ flavor symmetry on the $S^2$ supported by the co-dimension two defect, 
The 3d superconformal index in the presence of magnetic fluxes on two-sphere has been computed in the literature~\cite{Kim:2009wb,Imamura:2011su}. We can promote the computation to the 3d/5d coupled system.

The magnetic flux effectively shifts the flavor fugacity (or gauge fugacity when gauged) by powers of $q$ in the index, which implements the shifts in the angular momenta of the BPS modes when they couple to the background magnetic field.
In the localization computation, one can factorize the one-loop contribution into the product of two hemisphere partition functions on $S^1\times \mathbb{C}$ corresponding to the fixed point contributions at the North and South poles on two-sphere.
The scalar field in the flavor vectormultiplet takes nonzero vev proportional to magnetic flux $m$.
The flavor fugacity $\mu$ is combined with the scalar field and give a complex fugacity which becomes $\mu q^{m/2}$ at the north pole and $\mu^{-1} q^{m/2}$ at the south pole respectively~\cite{Benini:2013yva}.
We expect that the same shift would realize the inclusion of the background magnetic flux even when the 5d bulk coupling is considered.
We shall replace the flavor fugacity in the 5d perturbative part and the instanton part by the shifted fugacity.

Let us first consider the $D_{1,0}^{(1)}$ defect. We could turn on the background magnetic flux for the $U(1)$ flavor symmetry corresponding to the parameter $\mu = \sqrt{\lambda_1/\lambda_2}$. We would expect the superconformal index in the presence of the background monopole of charge $m$:
\begin{align}
&I_{1,0}(\lambda_1,\lambda_2,m;p,q) \\
&= I_{\mathrm{vec}}\oint \frac{da}{2\pi ia} \Big[a^{m}p^{m/4}(a^\pm;p,q)_\infty(pqa^\pm;p,q)_\infty I_{\mathrm{3dchiral}}(a;q) \left|Z^{1,0}_{\mathrm{inst}}(a,\lambda_1q^{m/2},\lambda_2q^{-m/2})\right|^2  + (a\rightarrow a^{-1})\Big] \ . \nonumber
\end{align}
Remember that the flavor fugacity $\mu$ is shifted by $q^{m/2}$ at the North pole while shifted by $q^{-m/2}$ at the South pole, implying that $(Z^{1,0}_{\mathrm{inst}})^* = Z^{1,0}_{\mathrm{inst}}(a^{-1},\lambda_1^{-1}q^{m/2},\lambda_2^{-1}q^{-m/2})$. The powers $a^{m}p^{m/4}$ in the integrand encode charge shifts of the vacuum in the monopole background. Such factors are ubiquitous in 3d index calculations, and  presumably could be determined by a precise localization in the 3d/5d system. Here we guess them in such a way to obtain consistent formulae. 

We now define two commuting sets of position and momentum variables:
\begin{equation}
	\hat{x}_\pm = q^{-m/2}\mu^{\pm} \,, \quad \hat{p}_\pm = e^{-\partial_m \pm \frac{1}{2}\log{q}\mu\partial_\mu} \ .
\end{equation}
They obey the commutation relations $\hat{p}_+\hat{x}_+ = q \hat{x}_+\hat{p}_+$ and $\hat{p}_-\hat{x}_- = q \hat{x}_-\hat{p}_-$.
The superconformal index $I_{1,0}$ is annihilated by following difference equations:
\begin{align}
	p^{-1/4}(\sqrt{p}\hat{p}_+ + \hat{p}_+^{-1})-\sqrt{\lambda}^{-1}(\hat{x}_+ +\sqrt{p}\hat{x}_+^{-1}) &= W(m)_S \ ,\nonumber \\
	p^{-1/4}(\sqrt{p}\hat{p}_- + \hat{p}_-^{-1})-\sqrt{\lambda}(\hat{x}_-+\sqrt{p}\hat{x}_-^{-1}) &= W(m)_N \ ,
\end{align}
where $W(m)_{S,N}$ are the indices of a fundamental Wilson loop inserted at the South and North pole respectively in the presence of magnetic flux $m$, normalized by the bare index at flux $m$.
When we evaluate the Wilson loop indices, the shift of the parameter $\mu$ should be properly considered: $\mu\rightarrow \mu q^{m/2}$ at the North pole and $\mu\rightarrow \mu q^{-m/2}$ at the South pole.

Numerical check for these equations can be done at any fixed flux $m$, order by order in $q$ and $p$ series expansion.
For example, we find the indices
\begin{align}
I_{1,0}(-1) &= p^{1/4}\sqrt{q}(\sqrt{\lambda}\mu+(\sqrt{\lambda}\mu)^{-1}) + p^{3/4}q^{3/2}(\lambda^{3/2}\mu+(\lambda^{3/2}\mu)^{-1} ) +\cdots \cr
I_{1,0}(1) &= p^{1/4}\sqrt{q}(\sqrt{\lambda}/\mu+\mu/\sqrt{\lambda}) + p^{3/4}q^{3/2}(\lambda^{3/2}/\mu+\mu/\lambda^{3/2} ) +\cdots
\end{align}
and the Wilson loop indices
\begin{align}
W(0)_S I_{1,0}(0)&= q(\sqrt{\lambda}/\mu+\mu/\sqrt\lambda)+\sqrt{p}q(\sqrt\lambda\mu)^{-1} - pq(\sqrt\lambda/\mu+\mu/\sqrt\lambda) + \cr
&\quad -\sqrt{p}q(p+q-q\lambda^2+q\lambda\mu^2)(\sqrt{\lambda}\mu)^{-1}+\cdots \ ,\nonumber \\
W(0)_N I_{1,0}(0)&= q(\sqrt{\lambda}/\mu+\mu/\sqrt\lambda)+\sqrt{p}q(\sqrt\lambda\mu) - pq(\sqrt\lambda/\mu+\mu/\sqrt\lambda) + \cr
&\quad -\sqrt{p}q(p\lambda^2\mu^2+q\lambda-q\mu^2+q\lambda^2\mu^2)(\lambda^{3/2}\mu)^{-1}+\cdots \ .
\end{align}
where the arguments of the indices denote magnetic fluxes $m$.
Plugging these indices one can check the difference equations at $m=0$ up to order $\mathcal{O}(p^{n_1}q^{n_2})$ with $n_1+n_2<2$.

\section{The equivariant index of a $SU(N)$ 5d gauge theory in the presence of 3d chiral fields} \label{sec:equiv}
In this section, we compute hemisphere partition function of the 5d guage theory on the $\Omega$-deformed $S^1 \times \mathbb{C}^2$ coupled to 3d chiral fields living on  $S^1\times \mathbb{C}$ sub-manifold. The superconformal index of the UV CFT can be expressed as a square modulus of these hemisphere partition functions. Computation of the partition function could be performed using supersymmetric localization. We refer the reader to \cite{Nekrasov:2002qd,Kim:2012gu} for the detailed explanation of localization of the 5d gauge theories without 3d fields. Here we shall concentrate on how the insertion of 3d fields affects the partition function computation.

We consider 5d $SU(N)$ gauge theory in the presence of 3d chiral multiplets which are in the fundamental representation of the bulk $SU(N)$ gauge group.
Localization reduces the partition function to an integral over saddle points which are given by moduli space of self-dual instantons on $\mathbb{C}^2$.

In order to calculate the contributions at the instanton saddle points one can start with the equivariant index of vector bundles over the instanton moduli space. We will work with the $U(N)$ instanton formulae, but we expect that the $SU(N)$ instanton results would be the same for the cases in this paper.
We propose that the 3d chiral multiplets introduce a three-dimensional vector bundle associated to the universal vector bundle $\mathcal{E}$ over the instanton moduli space. We then need to compute the equivariant index of the Dirac operator acting on the section of the vector bundle restricted on $\mathbb{C}$ in the fundamental representation of the $SU(N)$ gauge group. The result can be written in terms of the equivariant Chern character $Ch_{\mathcal{E}}$ of the universal bundle given in the appendix \ref{appendix:Nekrasov}:
\begin{equation}
\chi^{3d} = \frac{\mu_{3d}\sqrt{pq}}{(1-q)} \, Ch_{\mathcal{E}} = \frac{\mu_{3d}\sqrt{pq}\sum_{i=1}^Nz_i}{(1-q)}-\mu_{3d}(1-p)\sum_{I=1}^k\rho_I \ ,
\end{equation}
with $\mu_{3d}$, the equivariant parameter for the $U(1)$ flavor symmetry of the 3d fields. The factor $\frac{1}{1-q}$ reflects that the vector bundle is on the spatial sub-manifold $\mathbb{C}$ where $q$ is the equivariant parameter for the corresponding $SO(2)$ rotation group. The first term in the r.h.s is independent of the instanton number $k$ and thus it is interpreted as the perturbative contribution, while the other terms are interpreted as the non-perturbative contribution at the $k$ instanton saddle point.

Let us now compute the contribution to the partition function by the 3d matter fields. It can easily read off from the equivariant index $\chi^{3d}$. The equivariant index can be written as a sum over weights $\omega_a$ under the Cartan subalgebra of $SU(N)\times U(k) \times U(1)_p \times U(1)_q \times U(1)_f$ rotations where $SU(N)$ is the 5d gauge symmetry, $U(k)$ is the dual gauge symmetry in the ADHM construction, $U(1)_p$ and $U(1)_q$ are the spatial Lorentz rotation on $\mathbb{C}^2$, and $U(1)_f$ is a flavor symmetry: 
\begin{equation}
	\chi= \sum_{\omega_a} e^{\omega_a\xi_a}  \quad (e^{\xi_a}\in \{z,\rho,\mu_{3d},p,q\}) \ .
\end{equation}
The perturbative contribution in $\chi^{3d}$ has a factor $\frac{1}{1-q}$ and it should be understood as a power series expansion by $q$.
Applying the conversion formulae (\ref{eq:conversion}), we get the contribution of the 3d chiral multiplets to the perturbative partition function:
\begin{equation}
Z^{3d}_{\mathrm{pert}}(\mu_{3d},z_i,p;q) \sim \prod_{i=1}^N(\sqrt{pq}\mu_{3d}z_i;q)_\infty^{-1} \ .
\end{equation}
This is precisely the 3d partition function on $S^1\times \mathbb{C}$ of the free chiral doublets with fugacity $z_i$ for the $SU(2)$ flavor symmetry and $\sqrt{p}\mu_{3d}$ for the $U(1)$ flavor symmetry.
The symbol `$\sim$' denotes that the prefactor coming from regularization of the infinite product is ignored. The prefactor will become trivial anyway when we compute the superconformal index. 

Similarly, the contribution from the 3d chiral fields to the instanton partition function can be read off from the index $\chi^{3d}$. At $k$-instantons we get
\begin{equation}
Z_{k}^{3d}(\mu_{3d},\rho_I;q) = p^{\frac{k}{2}} \prod_{I=1}^k \frac{1-\mu_{3d}\rho_I}{1-p\mu_{3d}\rho_I}
\end{equation}
This 3d contribution is to be incorporated into the contour integral expression of the instanton partition function.
The full instanton partition function of the 3d/5d coupled system becomes
\begin{equation}
Z_{\mathrm{inst}}^{SU(N)}(z_i,\mu_{3d},\lambda;p,q) = \sum_k \frac{\lambda^k}{k!}  \oint \prod_{I=1}^k\frac{d\rho_I}{2\pi i \rho_I} \, Z_k^{5d}(z_i,\rho_I;p,q) \, Z_k^{3d}(\mu_{3d},\rho_I;q) \ .
\end{equation}
$Z_k^{5d}$ is the contribution from the 5d vectormultiplet given in (\ref{eq:instanton-pure}). We conclude that the full hemisphere index is given by
\begin{equation}
Z^{SU(N)}_{3d/5d}(z_i,\mu_{3d},\lambda;p,q) = (pq;p,q)_\infty^{N-1}\prod_{i\neq j}^N(pqz_i/z_j;p,q)_\infty Z^{3d}_{\mathrm{pert}}\, Z_{\mathrm{inst}}^{SU(N)} \ .
\end{equation}

We expect that the same instanton partition function would be computed from the partition function of $\mathcal{N}=(0,2)$ supersymmetric gauged quantum mechanics with the ADHM fields. Coupling to the 3d fields introduces additional matters, one fermi and one chiral multiplets, in the fundamental representation of the gauge group $U(k)$ on the quantum mechanics, which can be read off from $Z_k^{3d}$.

Let us now compute the contour integral over $\phi$ (or $\rho$).
We shall employ the Jeffrey-Kirwan (JK) residue prescription introduced in \cite{1993alg.geom..7001J,Benini:2013xpa}. It is known that the JK prescription also works for the Witten index computation in quantum mechanics. We will briefly review it now. See \cite{Hwang:2014uwa,Cordova:2014oxa,Hori:2014tda} for more detailed explanations.

The poles in the contour integral can be classified in terms of the charge $Q_i$'s of the multiplets contributing to the partition function.
Let us define hyperplanes in the $\phi$ planes where the integrand becomes singular.
Each charge vector $Q_i \in \mathbb{R}^k$ defines a hyperplane such as
\begin{equation}
	H_i = \{ \phi \in \mathbb{C}^k \, \big| \, Q_i(\phi) + z = 0\} \ ,
\end{equation}
where $z$ denotes other chemical potentials (or log of fugacities).
When $n\ge k$ hyperplanes intersect at a single point $\phi=\phi^*$ we can compute the residue around the point using the JK prescription. A set of charge vectors for the hyperplanes crossing the point is denoted by $\mathrm{Q}(\phi^*) \equiv \{Q_i \big| \phi\in H_i\}$.

We expand the integrand around each singular point $\phi^*$ in negative powers of $Q_i(\phi-\phi^*)$. Then the JK residue receives nontrivial contribution from the term of the form
\begin{equation}
\frac{1}{Q_{i_1}(\phi-\phi^*)\cdots Q_{i_k}(\phi-\phi^*)} \ ,
\end{equation}
where $Q_{i_1},\cdots,Q_{i_k}$ are $k$ charge vectors in $\mathrm{Q}(\phi^*)$. 

The JK prescription refers to a reference vector $\eta$ which can be arbitrary chosen in $\mathbb{R}^k$, but the final result is independent of the choice.
Given $\eta$, the JK residue is defined as follows \cite{1993alg.geom..7001J}:
\begin{equation}
\text{JK-Res}_{\phi^*}(\mathrm{Q}_*,\eta)\frac{d^k\phi}{Q_{i_1}(\phi)\cdots Q_{i_k}(\phi)}   = 
\left\{ \begin{array}{cl} \left|{\rm det}(Q_{i_1},\cdots, Q_{i_k})\right|^{-1} & \ \text{if} \ \eta\in \text{Cone}(Q_{i_1},\cdots,Q_{i_k}) \\ 0 & \ \text{otherwise} \end{array}\right.
\end{equation}
where `Cone' denotes the cone formed by the $k$ independent $Q_i$'s. We note that our $Z_k$ meets the projective condition which is required for the JK prescription \cite{1993alg.geom..7001J}: all poles in our $Z_k$ are non-degenerate. 
Applying this prescription the partition function can be written as
\begin{equation}
Z_k  = \frac{1}{k!}\sum_{\phi_*} \text{JK-Res} (Q_*,\eta) \ Z_k^{5d}(\phi,z)Z_k^{3d}(\phi,z) \ .
\end{equation}

Before introducing the 3d fields, this procedure reproduces the Young tableau sum expression of the instanton partition function \cite{Hwang:2014uwa}. We expect that the JK prescription works also for 3d/5d coupled systems.
The example in this section would provide a nontrivial evidence for that.

Let us look at a simplest example. At $k=1$ instanton, the instanton partition function with the 3d chiral fields is given by
\begin{equation}
Z_{k=1} = \oint \frac{d\phi}{2\pi i} \frac{1}{\prod_{i=1}^N4\sinh\frac{\phi-a_i+\frac{\epsilon_1+\epsilon_2}{2}}{2}\sinh\frac{-\phi+a_i+\frac{\epsilon_1+\epsilon_2}{2}}{2}}\cdot \frac{\sinh\frac{\phi+m}{2}}{\sinh\frac{\phi+m-\epsilon_1}{2}} \ ,
\end{equation}
where $\rho_1 \equiv e^{\phi}, z_i \equiv e^{a_i}, \mu_{3d} = e^m, p\equiv e^{-\epsilon_1}, q\equiv e^{-\epsilon_2}$.
For the rank one theory, the JK prescription says that we should pick up all the poles from the fields of positive charges. Such poles are from the factors of the form $(\sinh\frac{Q_i\phi+z}{2})^{-1}$ with $Q_i>0$. Thus the poles to be kept are
\begin{equation}
\rho_1 = z_i\sqrt{pq}\,, \quad \rho_1 = \mu_{3d}^{-1}p^{-1} \ .
\end{equation}
The latter is a novel pole coming from the 3d contribution and it is important to involve the residue at this pole in the instanton calculus. The sum over all residues gives the $k=1$ instanton partition function.
One can use the same prescription for the higher instanton computation.

As a check for our result, let us now compare the above partition function at $N=2$ with the hemisphere index of the 5d $SU(2)$ gauge theory in the presence of the $D_{0,1}^{(1)}$ defect at one of two fixed points, which was obtained by Higgsing the SQCD with $N_c=3$ and $N_f=2$.
Indeed, two partition functions are identical if we properly identify the parameters of two partition functions. More precisely, we find the relation
\begin{equation}
	II_{0,1}(a,\mu,\lambda;p,q) = Z_{3d/5d}^{SU(2)}(z_i,\mu_{3d},\lambda;p,q) \ ,
\end{equation}
upon the identification 
\begin{equation}
	z_1 = 1/z_2 = \sqrt{a}\,, \quad \mu_{3d} = \mu q^{-1/2} p^{-3/4} \ .
\end{equation}
The perturbative parts trivially match. For the instanton parts we have checked the equivalence of two partition functions till $k=2$.

\section{Conclusions and open questions}\label{sec:conc}
Our first conclusion is that the Higgsing prescription provides an effective methods to compute supersymmetric indices and partition functions 
in the presence of a large variety of co-dimension two defects. When the bulk field theories admit a brane construction which makes the Higgs branch manifest, the defects also admit such construction. The answer often takes the form of a sum over fixed points 
of some moduli space of position-dependent Higgs branch configurations, which appear to have a direct interpretation in the brane language. 

Our second conclusion is that the index of a 3d/5d system, and thus presumably the $S^4_b$ or instanton partition function of 
a 2d/4d system, transforms just as a 3d index ($S^2$ partition function) under basic 3d (2d) operations involving the defect degrees of freedom, such as adding extra chiral matter fields and gauge fields living in co-dimension two and coupled to the original defect through 
defect superpotential terms or gauging defect flavor symmetries respectively. Indeed, in this paper we subjected the 3d/5d indices 
to such transformations and obtained results compatible with the dualities expected from the brane pictures. 
In particular, this means that we can combine the Higgsing construction and such manipulations to greatly extend the class of ``computable''
co-dimension two defects. 

Our third conclusion is that the index of five-dimensional SCFTs in the presence of co-dimension two defects 
should satisfy difference equations which quantize the Seiberg-Witten curve, but involve the insertion of 
bulk line defects in the index, and thus do not form a finite closed system of equations. 
We expect to recover closed systems of difference equations in the NS limit.

Our work leaves several open questions
\begin{itemize}
\item The Higgsing prescription and the properties of the index under coupling extra degrees of freedom in co-dimension two should be 
tested more systematically in five dimensions. An obvious example is to test the non-Abelian $S$ transformations and dualities for 
defects associated to multiple parallel D3 branes. A set of $N$ D3 branes ending on $N$ auxiliary D5 five-branes (``Dirichlet defect'')
should possess a $U(N)$ flavor symmetry on the defect. Coupling such flavor symmetry to 3d triangular quiver gauge theories 
with $U(N)$ flavor symmetry such as $T[U(N)]$ \cite{Gaiotto:2008ak} by gauging the 3d diagonal $U(N)$ flavor symmetry should 
reproduce defects where the D3 branes end on auxiliary NS5 branes. 

A second example would be to consider Higgs branches where two non-trivial webs are separated from each other.
The defect associated to $N$ D3 branes in such a system should arise from gauging the diagonal $U(N)$ 
symmetry of the Dirichlet defects of the two sub-theories. 

\item The Higgsing prescription and the difference equations satisfied by the defects for linear quiver gauge theories 
should admit an interpretation in the language of $q$-deformed Virasoro and W-algebras, see. e.g. \cite{Awata:2009ur,Awata:2010yy,Nieri:2013yra,Nieri:2013vba,Taki:2014fva,Mitev:2014isa}. 

\item Connections between various gauge theories and integrable systems have been observed in the literature. See~\cite{Cordova:2014oxa} and references therein (see also~\cite{Nekrasov:2009uh}).  We have noticed that our partition functions in the NS limit are related to the eigenfunctions of the two-body closed q-Toda system and their difference equations are the quantization of the integrable Hamiltonian. The bi-spectral duality of the integrable systems is realized as the duality of the brane diagrams.
It would be natural to consider generalization of these connections to more complicated systems. Our work may provide a framework to compute eigenfunctions of hamiltonians of certain integrable models.
Also, the chain of dualities in the brane pictures may provide new bi-spectral dualities and integrable systems.
The generalization to the 5d $\mathcal{N}=1^*$ theories coupled to 3d defects and their cousins will be the subject of a forthcoming publication~\cite{BKK2014}.

\item We do not know under which conditions the defects obtained from Higgsing admit a direct definition in the IR theory
in terms of coupling the bulk theory to degrees of freedom living in co-dimension two. In five dimensions the question is somewhat ill-defined: 
the gauge theory description is not UV complete, and thus one can at most seek low-energy effective descriptions of a defect defined in the UV by Higgsing a more complicated SCFT. 

In four dimension the question is meaningful, and the answer somewhat mysterious. For many defects, say 
the defects associated to D2 branes ending on auxiliary NS5 branes, the brane construction provides such a direct description as a 2d gauge theory coupled to the bulk gauge degrees of freedom. The index or partition function can be reproduced directly from such a description \cite{Gadde:2013dda}. 

For some other defects though, such as these associated to D2 branes ending on 
auxiliary D4 branes, the situation is confusing. The D2 brane has a moduli space 
which explores the neighbourhood of all NS5 branes, each giving a distinct description in terms of 2d degrees of freedom 
 coupled to the bulk theory. From the point of view of quantities such as the effective low-energy twisted superpotential  
 each individual description seems to be sufficient to recover the whole moduli space \cite{Gaiotto:2013sma}, 
 by analytic continuation through strongly-coupled values of the parameters. Our computation of the index, though, 
 involves a {\it sum} over all these contributions, and an individual description would likely fail to reproduce the full answer. 
 This issue deserves further investigation. 

\item It would be interesting to figure out which gauge theory boundary conditions have a UV completion. The hemisphere indices may help answer that question. 

\end{itemize}

\section*{Acknowledgements}

We are grateful to J. Gomis, M. Bullimore, P. Koroteev, S. Kim for useful discussions. HC would like to thank the organizers of ``Exact Results in SUSY Gauge Theories in Various Dimensions'' at CERN and also CERN-Korea Theory Collaboration funded by National Research Foundation (Korea) for the hospitality and support.
The research of DG and HK was supported by the Perimeter Institute for Theoretical Physics. Research at Perimeter Institute is supported by the Government of Canada through Industry Canada and by the Province of Ontario through the Ministry of Economic Development and Innovation.

\appendix

\section{Nekrasov partition function}\label{appendix:Nekrasov}
The Nekrasov partition function can be computed using the equivariant localization on the moduli space of self-dual instantons in 5d (or 4d) gauge theories~\cite{Nekrasov:2002qd,Nekrasov:2003rj}. We shall focus on 5d $\mathcal{N}=1$ SYM with $U(N)$ gauge group. The moduli space of the self-dual instantons can be described by the ADHM data subject to the ADHM constraints. 

Let us briefly review the ADHM construction. For $k$ instantons, we have two vector spaces $V$ and $W$ with complex dimensions ${\rm dim}_{\mathbb{C}}\, V = k$ and ${\rm dim}_{\mathbb{C}}\, W = N$. The ADHM data associated to the vector spaces consist of the ADHM fields $A,B\in {\rm Hom}(V,V),\, P \in {\rm Hom} (W,V) ,\, Q \in {\rm Hom} (V,W)$. We can construct the moduli space of $k$ instantons using the following hyper-K\"ahler quotient
\begin{equation}
    \mathcal{M}_{N,k} = \{(A,B,P,Q)|\mu_{\mathbb{C}} = 0 \} / {\rm GL}(k,\mathbb{C})
\end{equation}
with the ADHM equation 
\begin{equation}
    \mu_{\mathbb{C}} := [A,B] + PQ = 0 \ .
\end{equation}

Let us first compute the equivariant Chern characters and indices for vector bundles on the instanton moduli space from which we can easily compute the partition function.
The equivariant character for the universal bundle $\mathcal{E}$ is given by~\cite{Losev:2003py,Shadchin:2004yx}
\begin{equation}
Ch_{\mathcal{E}}(a,\phi;\epsilon_1,\epsilon_2) = \sum_{i=1}^Ne^{a_i} - (1-e^{-\epsilon_1})(1-e^{-\epsilon_2})e^{\epsilon_+}\sum_{I=1}^ke^{\phi_I} \ .
\end{equation}
$\epsilon_1,\epsilon_2$ are the equivariant parameters for the rotations on $\mathbb{C}^2$ and $\epsilon_+\equiv(\epsilon_1+\epsilon_2)/2$. $a_i$ and $\phi_I$ are the equivariant parameters for the Cartans in $U(N)$ and $U(k)$ gauge groups respectively. Using this character we compute the equivariant index of the tangent bundle $\mathcal{TM}$ over the instanton moduli space:
\begin{align}
	\mathrm{ind}_{\mathcal{TM}} &= -\frac{Ch_{\mathcal{E}}\ Ch_{\mathcal{E^*}}}{(1-e^{-\epsilon_1})(1-e^{-\epsilon_2})} \\
	&= -\frac{\sum_{i,j=1}^N e^{a_{ij}}}{(1-e^{-\epsilon_1})(1-e^{-\epsilon_2})} + e^{\epsilon_+}\sum_{i=1}^N\sum_{I=1}^k (e^{\phi_I-a_i}+e^{a_i-\phi_I}) - (1-e^{\epsilon_1})(1-e^{\epsilon_2})\sum_{I,J=1}^ke^{\phi_{IJ}} \ ,\nonumber
\end{align}
with shorthand notations $a_{ij}=a_i-a_j$ and $\phi_{IJ}=\phi_I-\phi_J$.
The first term in the second line is independent of the instanton number $k$ implying that it corresponds to the perturbative contribution of the $U(N)$ vectormultiplet. The denominator factor should be understood as a power series expansion with respect to $e^{-\epsilon_1},e^{-\epsilon_2}$ (or $e^{\epsilon_1},e^{\epsilon_2}$ depending on the orientation).
The other terms are the $k$ instanton contributions.

One can consider fundamental hypermultiplets in the 5d gauge theory. They introduce additional fermion zero modes to the instanton moduli space. The equivariant index for a fundamental hypermultiplet is given by
\begin{equation}
{\rm ind}_{\rm fund} = e^{m-\epsilon_+}\frac{Ch_{\mathcal{E}}}{(1-e^{-\epsilon_1})(1-e^{-\epsilon_2})} = \frac{e^{m-\epsilon_+}\sum_{i=1}^Ne^{a_i}}{(1-e^{-\epsilon_1})(1-e^{-\epsilon_2})} - \sum_{I=1}^ke^{\phi_I+m} \ ,
\end{equation}
where $m$ is the equivariant parameter for the $U(1)$ flavor symmetry.
It is then straightforward to compute the partition function using the conversion formula:
\begin{equation}\label{eq:conversion}
 \mathrm{ind} = \sum_{i}n_ie^{z_i} \ \rightarrow \ \prod_i \big(2\sinh\frac{z_i}{2}\big)^{-n_i} \ .
\end{equation}
The hyperbolic sine factor reflects that the temporal circle is fibered over the moduli space.
The 5d Nekrasov instanton partition function for the $\mathcal{N}=1$ $U(N)$ gauge theory is then given by the following integral expression:
\begin{align}\label{eq:instanton-pure}
	Z_{\rm inst}^{U(N)} &= \sum_{k=0}^\infty \lambda^k Z_k \,, \\
	Z_k &=\frac{1}{2^kk!} \oint \prod_{I=1}^k \left[\frac{d\phi_I}{2\pi i} \prod_{i=1}^N\frac{1}{4\sinh\frac{\phi_I-a_i+\epsilon_+}{2}\sinh\frac{-\phi_I+a_i+\epsilon_+}{2}}\right] \prod_{I,J=1}^k \frac{\sinh'\frac{\phi_{IJ}}{2}\sinh\frac{\phi_{IJ}+2\epsilon_+}{2}}{\sinh\frac{\phi_{IJ}+\epsilon_1}{2}\sinh\frac{\phi_{IJ}+\epsilon_2}{2}} \ , \nonumber
\end{align}
with $Z_{k=0}=1$. The prime in the hyperbolic sine indicates that $\sinh(0)$'s are omitted. When the theory has $N_f$ fundamental hypermultiplets we need to multiply the following contribution to the integrand:
\begin{equation}
	Z_k^{N_f}(\phi_I,m_a) = \prod_{I=1}^k\prod_{a=1}^{N_f} 2\sinh\frac{\phi_I + m_a}{2} \ .
\end{equation}

The integral can be evaluated by choosing unit circle contours for $\phi_I$'s and assuming $e^{-\epsilon_2} < e^{-\epsilon_1} \ll 1$ or by using the JK-residue prescription. It turns out that the poles are classified by the $N$-tuple of Young tableaux $\vec{Y}=\{Y_1,Y_2,\cdots,Y_N\}$.
Each Young tableau contains $k_i$ boxes and the total number of boxes is $k=\sum_{i=1}^Nk_i$. Denoting by $(m,n)$ the position in the $i$-th Young tableau, the corresponding pole is given by
\begin{equation}
\phi(m,n) = a_i + \epsilon_+ - m\epsilon_1 - n\epsilon_2 \ .
\end{equation}
The residue at the pole yields
\begin{equation}
	Z_{\vec{Y}}(a_i,m_a;\epsilon_1,\epsilon_2) = \prod_{i=1}^N\prod_{(m,n)\in Y_i} \frac{\prod_{a=1}^{N_f}\sinh\frac{\phi(m,n)+m_a}{2}}{\prod_{j=1}^N 4\sinh\frac{E_{ij}(m,n)}{2}\sinh\frac{E_{ij}(m,n)-2\epsilon_+}{2}} \ ,
\end{equation}
where
\begin{equation}
E(m,n) = a_i-a_j - \epsilon_1 h_i(m,n)+\epsilon_2(v_j(m,n)+1) \ .
\end{equation}
$h_i(m,n)$ is the horizontal distance from $(m,n)$ to the end of $m$-th row of the $Y_i$. $v_j(m,n)$ is the vertical distance from $(m,n)$ to the end of $n$-th column of the $Y_j$. Then the full partition function at $k$ instantons is the sum of the residues:
\begin{equation}
Z_k(a_i,m_a;\epsilon_1,\epsilon_2) = \sum_{\sum_{i=1}^N k_i=k} Z_{\vec{Y}}(a_i,m_a;\epsilon_1,\epsilon_2) \ .
\end{equation}

The perturbative contribution reads from the equivariant indices:
\begin{equation}
Z_{\rm pert}(a_i,m_a;\epsilon_1,\epsilon_2) = \prod_{m,n=0}^\infty\frac{\prod_{i,j=1}^N2\sinh'\frac{a_{ij}-m\epsilon_1-n\epsilon_2}{2}}{ \prod_{i=1}^N\prod_{a=1}^{N_f}2\sinh\frac{a_i+m_a-\epsilon_+ -m\epsilon_1-n\epsilon_2}{2}} \ .
\end{equation}

\bibliographystyle{JHEP}

\bibliography{5d-paper}

\end{document}